\documentclass[preprint,aps,amsmath,amssymb,showpacs,naturemag]{revtex4-1}
\usepackage{amsfonts}
\usepackage{amssymb}
\usepackage{amsmath}
\usepackage{notes2bib}
\usepackage{graphicx}
\usepackage{soul}
\usepackage[table,xcdraw]{xcolor}
\usepackage{dcolumn}
\usepackage{epstopdf}
\usepackage[export]{adjustbox}
\usepackage{bm}
\usepackage[T1]{fontenc}
\usepackage{hyperref}
\usepackage{mwe}
\usepackage{tabularx}
\usepackage{wasysym}
\usepackage[final]{pdfpages}
\usepackage{etoolbox} 

\makeatletter
\patchcmd{\@outputpage@head}{\@ifx{\LS@rot\@undefined}{}{\LS@rot}}{}{}{}
\makeatother

\definecolor{Gray}{gray}{0.89}

\begin{document}

\title{Unconventional fractional quantum Hall states and Wigner crystallization in suspended Corbino graphene}

\author{Manohar Kumar$^{1\ast \dagger}$, Antti Laittinen$^{1}$, Pertti Hakonen$^{1}$}
\email{To whom correspondence should be addressed; pertti.hakonen@aalto.fi.}
\affiliation{$^1$Low Temperature Laboratory, Department of Applied Physics, Aalto University, Espoo, Finland}
\affiliation{$^{\dagger}$Laboratoire Pierre Aigrain, D\'{e}partement de Physique de l'\'{E}cole Normale sup\'{e}rieure, Paris, France}


\begin{abstract}

Competition between liquid and solid states in two-dimensional electron system is an intriguing problem in condensed matter physics. We have investigated competing Wigner crystal and fractional quantum Hall ( FQH ) liquid phases in atomically thin suspended graphene devices in Corbino geometry. Low temperature magnetoconductance and transconductance  measurements along with $IV$ characteristics all indicate strong charge density dependent modulation of electron transport. Our results show unconventional FQH phases which do not fit the standard Jain's series for conventional FQH states, instead they appear to originate from residual interactions of composite fermions in partially filled higher Landau levels. And at very low charge density with filling factors $\nu \lesssim$ 1/5,  electrons crystallize into an ordered Wigner solid which eventually transforms into an incompressible Hall liquid at filling factors around $\nu \leq$ 1/7. Building on the unique sample structure, our experiments pave the way for enhanced understanding of the ordered phases of interacting electrons.

\end{abstract}


\maketitle

\section*{Introduction}
In the extreme quantum limit, the interplay between kinetic and potential energy of electrons in the two dimensional electron gas leads to formation of exotic states like fractional quantum Hall (FQH) state, a many body state where elementary excitations have fractional electronic charge \cite{Tsui1982}, or even formation of electron solids {\it i.e.} Wigner crystals where electrons freeze into a periodic lattice at very low temperature \cite{Wigner1934}. The FQH states are incompressible liquid states with quantized Hall conductance at distinct fractional fillings $\nu = n\phi_0/B$, specified by integer values $\nu_{CF}, m$ = 1, 2... according to $\nu = \nu_{CF}/(2m\nu_{CF}\pm 1)$, at charge density $n$ and magnetic field $B$ \cite{Jain1989}. Here $\phi_0 = h/e$ is the flux quantum. In the presence of strong Coulomb interactions, electrons may eventually condense into quasiparticles of unconventional fractional states $\nu$ at which $\nu_{CF}$ attains non-integer values~\cite{Pan2003}. Some of these unconventional FQH states are of special interest due to the non-Abelian properties of their quasiparticles~\cite{Mukherjee2012}. The manifestation of these exotic states requires a large Coulomb interaction energy compared with the disorder potential, putting strict  requirements for the quality of the two-dimensional electron gas (2-DEG), the strength of the magnetic field, and for excitation by environmental noise or temperature.

Owing to reduced screening in atomically thin graphene, the electrons interact with  higher Coulomb interaction energy than in conventional semiconductor heterostructures, providing an extraordinary  setting for both unconventional FQH states and Wigner crystals. However, to our knowledge there are yet no previous experimental results showing manifestation of these exotic states in graphene. So far, most of the quantum Hall studies in graphene have been performed using supported samples (G/h-BN) in Hall bar geometry. The multi-terminal magnetotransport measurements upto 35 T on these samples have revealed several FQH states of mostly integer multiples of 1/3 ({\it i.e. } 1/3, 2/3 \ldots13/3) ~\cite{Dean2011}. Contrary to this, despite of the higher Coulomb interaction energy in suspended graphene compared to supported samples and the very first observation of fractional state of $\nu$ = 1/3 in rectangular two terminal high-mobility suspended samples, not much progress has been seen in magnetotransport in suspended samples~\cite{Bolotin2009, Du2009}. It was realized that these difficulties are mostly due to either intermixing of  $\rho_{xx}$ and $\rho_{xy}$~\cite{Abanin2009} or strain-induced pseudomagnetic fields ~\cite{Prada2010} in two-terminal and multi-terminal suspended graphene samples. This limits the experimental verification of fractional states to either local compressibility measurements using SET techniques~\cite{Feldman2012} or transconductance measurements \cite{Lee2012}, both of which techniques have limited access to the complete electrical transport characteristics of fractional states; they probe local properties only. Despite of the experimental challenges, the higher Coulomb interaction energy in suspended graphene \cite{Nayak2014} makes it a very attractive platform to probe the interaction-based physics in conjunction with mechanical motion~\cite{Chen2016} and to investigate the competition between the lowest order FQH states and Wigner-crystal type of ordering \cite{Wigner1934}.

In this work, we have fabricated suspended graphene Corbino disks for probing correlated many-body physics in graphene. Our sample structure is illustrated in Fig. 1; for detailed fabrication steps and sample characterization see the supplemental information (Fig. S1 and S2).  The Corbino geometry outperforms the regular Hall bar geometry, as it directly probes the bulk two-dimensional electron gas (2-DEG) without having complications due to edge state transport \cite{Panchal2014}. In this geometry, the quantum Hall edge states counterpropagate at the perimeters of Corbino disk and no edge channel is formed between the two electrodes. Very recently, such quantum Hall states have been probed in graphene by magnetotransport measurement in Corbino geometry, also in on-substrate devices~\cite{Zhao2012,Peters2014}. These measurements with on-substrate graphene Corbino disks have failed to show any fractional states, perhaps due to strong charge inhomogeneity induced by the substrate.
Contrary to earlier approaches, in our experiment on suspended graphene Corbino disks, the quantum transport is dominated by activation or tunneling across  quantum states localized on static disorder potential. In suspended graphene, impurities and uncontrolled charge doping create a disorder potential forming localized states; which plays an important role in the magnetotransport. Like in a quantum Hall states, in  fractional quantum Hall regime, these states are formed to regions where the charge density does not lead to complete filling of a FQH state (Fig S6) \cite{Martin2004b}. An incompressible FQH liquid surrounds the localized states, inside which free rearrangement of charge leads to a constant electrochemical potential. Such edge states can carry current across the sample only via coupling, either by quantum tunneling or by thermal activation, to adjacent localized states. Quasi-particles in single units can be added into the localized states, and thus such states behave quite like quantum dots in semiconducting devices. In the case of weak screening, these localized states interact electrostatically with many nearby localized states. Charge fluctuations among the interacting quantum dots lead to broadening of the energy levels, which enhances probability of quantum tunneling across localized states at low temperature. The charging effects lead to sets of modulation patterns in the current $I(B,V_g)$ through the device as an AC gate voltage $V_g$ is applied. This is presented as transconductance maps $g_m(B,V_g) = dI/dV_g$ which facilitates identification of localized states by pattern recognition and their classification according to the dominant filling factors \cite{Lee2012}.

Here we used the transconductance measurement either by numerical derivative of logarithmic differential conductance $\delta log G_d/\delta n$, see Fig. 2A, or by lockin measurements of $g_m(B,V_g)$, see Fig. 2B, to identify the fragile fractional states. As seen in Fig. 2, both FQH and QH states, localized on disorder potential landscape, are observed as additional sets of parallel lines in our transconductance scans. The slope of the fringes varies in non-monotonic fashion, which is indicative of the features coming from spatially-modulated charge density due to disorder potential and corresponding to filling factor $\nu$. These fringes have been treated using a cross-correlation analysis described in the supplemental information (Fig S4), Fig. 2C.  A summary of the FQH states identified by $\delta \log(G_d)/\delta n$ and $g_m$ measurements in our four Corbino samples is displayed in Table 1. In our magnetoconductance and transconductance measurements, we resolved a distinctive set of incompressible liquid states with fractional filling factors of \{-1/3, 1/5, 2/7, 4/13, 1/3, 4/11, 2/5, 3/7, 4/9, 4/7, 3/5, 2/3, 4/5, 4/3\}. Of these FQH states, $\nu$ = 4/13 and 4/11 are unconventional states, formed due to interactions between composite fermions~\cite{Arkadiusz2004}. Our activation transport measurements on $\nu = 4/13$ and $\nu = 4/11$ states verify their incompressible nature with an energy gap of the order of $2\%$ of the gap energy at $\nu = 1/3$. In addition to our magnetoconductance measurements, {\it IV} characteristics and microwave spectroscopy show evidence of solid phase order of electrons, {\it i.e.} Wigner crystallization, at low densities around $\nu \simeq 1/5-1/7$~\cite{Platzman1993, Pan2002}.
\section*{Results}
\subsection*{Fractional quantum Hall states of composite fermions }
The majority of fractional states in two dimensional electron gases can be described using Laughlin's wave function and Jain's composite fermion (CF) theory. In the CF picture, fractional states are interpreted as integer quantum Hall effect of flux-transformed noninteracting composite particles. These CFs are quasiparticles of an electron bound to an even number of flux quanta in a effective magnetic field $B^* = B - 2\phi_0 n$. Even though Jain's theory of CF considers these particles as noninteracting entities, there may remain small interactions between the CFs which lead to the formation of exotic states {\it e.g.} fractional states of CFs~\cite{Arkadiusz2004}. These unconventional fractional quantum Hall states reside at fillings $\nu_{CF}^* = \nu_{CF}/ (2m\nu_{CF} \pm 1)$, where $\nu_{CF}$ has fractional non-integer values and the integer $m$ is the number of flux quanta \cite{Mukherjee2015}. Evidence of such interacting CF states has been obtained by Pan {\it et al.} in GaAs/AlGaAs \cite{Pan2003}. Despite of the stronger electron-electron Coulomb interactions, these unconventional fractional states of composite fermions have remained elusive in graphene. Our present data provides the first evidence of unconventional FQH states $\nu =$ \{4/13, 4/11\} in suspended graphene.

These states are distinctly seen as peaks in $G^{-1}_d$ at filling factors across $\nu = 1/3$, as depicted in Fig. 3A. The $G^{-1}_d$ vs $\nu$ plot for fields $B = 6 - 8.5$ T shows distinct peaks at $\nu = 4/13$ and $4/11$. Since these peaks seem to evolve at the specific filling factors without significant movement, we conclude that they are due to unconventional fractional states rather than replicas of other conventional FQH states shifted by local charge inhomogeneity. The $\nu$ = 4/11 corresponds to the fractional filling $\nu_{CF} = 1 + 1/3$ for two flux quanta composite fermion ($m$ = 1), and $\nu$ = 4/13 corresponds to $\nu_{CF} = 1 + 1/3$ for four flux quanta composite fermion ($m$ = 2)~\cite{Arkadiusz2004, Pan2015, Samkharadze2015}.

In the temperature dependence of magnetoconductance, the incompressibility of the FQH states shows up as an activated behavior in the DC conductivity $\sigma_{xx} = \sigma_0\exp(-\Delta_{\nu}/2k_B T)$. From the Arrhenius plot shown in Fig. 3B, the linear fit to $\log_{10}\sigma_{xx}$ vs $1/T$ yields the activation energy $\Delta_{\nu}$. The activation energy of $\nu$ = 1/3 state amounts to $\Delta_{1/3}$ = 4.7 K at 7 T, in agreement with Ghahari {\it et al.} \cite{Ghahari2011}. Similarly, the activation energies for $\nu$ = 4/13 : $\Delta _{4/13}$ 110 mK and $\nu = 4/11$: $\Delta _{4/11}$ = 150 mK. These energies are about 2 \% of the activation energy of the $\nu$ = 1/3 energy gap, in agreement with theoretical expectations~\cite{Mukherjee2015}.

\subsection*{Wigner Crystallization}
The FQH states at the lowest charge densities are of special interest; the correlated liquid states are expected to be terminated in an electron solid phase. The formation of this electron solid, {\it i.e.} the Wigner crystal, is related to two competing energy scales: the Coulomb potential energy and the kinetic energy. At a sufficiently low charge density and high magnetic field, the Coulomb energy dominates and, hence, electrons crystallize into a two dimensional electron solid.  In our experiments around $\nu$ = 1/7 - 1/5, we find evidence for such incompressible solid states within the lowest Landau level. The Landau fan diagram for $\nu < 1/5$ is shown in Fig. S6 a-b. We affirm the incompressibility of the region ($\nu$ = 1/7 - 1/5)  by microwave spectroscopy, activation transport, and  non-linear current-voltage characteristics.

Upon the onset of Wigner crystallization, electrons undergo a phase transition at which structural order and extent of correlations are enhanced. Thermodynamically, the incompressibility of a system scales up as it undergoes a phase transition from liquid to solid. We probed this increased incompressibility of electrons in the Wigner solid phase by performing microwave absorption spectroscopy. The measurement scheme is shown in the supplemental information (Fig S8). An rf signal of amplitude $-68$ dBm was fed from the microwave generator to the inner contact, and the DC conductance was simultaneously measured at a 50 $\mu$V bias voltage using a transimpedance amplifier. For the rf-power of -68 dBm, the mixing chamber temperature stayed within $20\pm10$ mK. However, assuming validity of the Wiedmann-Franz law, the minimum achievable electron temperature at this rf-power remains around 0.4 K according to the regular hot electron model.

Fig. 4A displays the measured dc current $I$ while sweeping the frequency of the rf signal over the range $0.01 - 5$ GHz. The data in Fig. 4A display a strong resonance at $f_p =$ 3.0 GHz for $\nu = 0.16$. This resonance is visible in the region 1/7 $< \nu <$ 1/5, while it becomes very weak at $\nu$ = 1/3 and fully absent at $\nu \leq$ 0.12. This resonance in microwave absorption spectra, observable up to $T\sim 1.4$ K, can be understood either as a resonance of individual localized particles or as a collective resonance of an in-compressible, ordered phase. For photon-induced carrier excitation at $f_p = 3$ GHz from localized states, the critical temperature for thermally-induced single carriers would be $T_I < hf_p/k_B \sim~ 140$ mK. Contrary to this, the absorption resonance is observed up to $T_m = 1.5$ K, exceeding $T_I$ by an order of magnitude. At such a high temperature, the single particle localization as formulated above is not effective and all particles would become delocalized. Hence, this peak in the microwave absorption spectra points towards the existence of an in-compressible solid as for its origin. In the presence of an external oscillating field, the electrons in the Wigner crystal domains oscillate within their pinning potential with a collective resonance frequency of $f_p$ \cite{Fukuyama1978}. Also, for the correlated Laughlin liquid, an rf-field will excite the lowest lying collective excitation, the magnetoplasmon of the FQH state.
The energy scale of magnetoplasmons ($E/\hbar  \sim 100$ GHz for our sample size) is much larger than that of transverse phonon-like modes in the Wigner crystal and, consequently, these plasmons remain beyond reach in our experimental setting \cite{Fukuyama1978, Fogler2000, Chitra2001}.

The sharpness of the low-temperature resonance spectra is at odds with the presence of complex non-uniform states, such as Wigner glass or non-homogeneous two-fluid states, which would tend to give a broader microwave resonance signal. The microwave absorption spectroscopy was also performed at two times larger rf-irradiation power ($P = -65$ dBm) in the Wigner crystal regime (0.153 $<\nu<$  0.172), see Fig. 4B. The resonance at $\nu = 0.16$ is the sharpest,  while away from $\nu = 0.16$, the resonances are either broader or of smaller amplitude.   At the rf-power of -65 dBm, the combined action of increased activated transport and enhanced Joule heating due to rf absorption leads to an increase in the dc current $I$, see also supplemental information (Fig. S9).

For further proof of the Wigner crystal, we also probed the $IV$ characteristics of our sample.  The non-linear electrical conductance provides a means to distinguish between several possible transport mechanisms, including activation of fractionally charged quasiparticles \cite{DAmbrumenil2011} and depinning of Wigner crystal \cite{Goldman1990, Williams1991}. The $IV$ characteristics and the relevant transport mechanisms at low filling factor ($\nu$ = 0.16 ... 0.33) are discussed in the supplemental information for sample EV (Fig. S7 a) . The behavior of our non-linear $IV$ characteristics for sample XD is illustrated in Fig. 4C on a semilog scale. The data at $\nu$ = 0.16 display exponential increase in the current $I$ between $V$ = 0.1 - 1.8 mV. Such exponential increase in current has been identified as gradual depinning of the ordered charge carrier phase,  having either charge density wave (CDW) or Wigner solid type of order; it is more likely to be a Wigner crystal order at low charge density in graphene \cite{Zhang2007, Knoester2016}. Deviation of exponential behavior starts at $V_T$ = 1.8 mV which is identified as the depinning threshold of Wigner crystal sliding. When $V < V_T$, charge transport takes place by thermally activated hopping of pinned regions. Similar behavior was observed in sample EV2, where we obtained a threshold voltage of $V_T$ = 1.4 mV.

Following Ref. \cite{Williams1991}, we write for the current due to thermally activated motion of pinned crystallites below $V_T $ as
\begin{equation}
I_W=e^*f_a \{\exp[-\frac{\bar{\Delta}-eV/2N}{k_B T}]-\exp[-\frac{\bar{\Delta}+eV/2N}{k_B T}]\}
\label{eq:depinning}
\end{equation}
where $e^*$ is the effective charge, $f_a$ is the depinning attempt frequency, $\bar{\Delta}$ is the average value of the pinning potential, and $N$ denotes the number of Wigner crystallites in series across the transport path. In Fig. 4C, we obtain a good agreement between this basic theory and our experimental data at $\nu = 0.16$ using parameters $e^* = e$, $f_a =$ 3.0 GHz, $\bar{\Delta}$ = 140 $\mu$eV, $N = 7$, and $T$ = 0.2 K. As expected for $N$ Wigner crystallites in series, $N\bar{ \Delta} /e \simeq V_T/2 = 1.0$ mV for EV2.

To elucidate the strength  of the pinning barrier, we probed the  melting temperature of the Wigner crystal by recording the amplitude of the pinning resonance of Fig. 4A as a function of $T$, which is presented in Fig. 4D. The increase in the electronic temperature due to the -68 dBm rf-excitation is much smaller than the melting temperature, but it may account for the saturation in $\Delta I$ at the lowest temperatures. Hence, the initial rise in the peak current at low temperatures is due to heating assisted activated transport upon rf irradiation, while at higher temperature, melting of the Wigner crystal results in a weak temperature dependence of $I(T)$ leading to a decrease in the peak amplitude \cite{Drichko2016}. Fig. 4D, shows $T_m \sim$ 1.5 K, which agrees closely with the fitted activation barrier $\bar{\Delta}$ = 170 $\mu$eV. Also, this $T_m$ correlates with the classical theoretical estimate of the melting temperature $T_{m} = e^2\sqrt{n}\/(4\pi\epsilon_o\epsilon_g k_B\Gamma) \simeq$1 K\cite{Chen2006}, where $\epsilon_g = 3$, is the dielectric constant for graphene and $\Gamma = 130$ \cite{Fertig1999}. Besides the non-quantum formula for shear modulus, this comparison is influenced by uncertainty due to non-uniformity and irregular melting of the Wigner crystal. Clearly, our microwave resonance data in Fig. 4B indicate the presence of crystallites which melt before the electronic temperature has reached the dominant $T_m$  shown in Fig. 4D.

In our rf-spectroscopy measurements, we investigated microwave-induced phenomena also around the pinning frequency $f_p = $ 3.0 GHz. In general, a  coherent crystallite averages over the pinning potential and the frequency becomes smaller with enhanced crystal quality. Consequently, an unavoidable spread in crystallite frequencies exists and it is expected to lead to a broad resonance, which is observed in our experiments under large rf-irradiation as illustrated in Fig. 4B.  We can estimate the average domain size of pinned Wigner crystallites $L_c$ = 0.70 $\mu$m from classical shear modulus $c$ and resonance frequency $f_p$\cite{Bonsall1977} (see supplemental information). In comparison to thermal depinning model, the size of the crystallites obtained here is significantly larger. The discrepancy could be due to an overestimation of $L_c$ by the classical-shear-modulus model where the renormalization of the traverse phonon dispersion relation by quantum effects has been ignored, although this is  disputable at very low charge density and in high magnetic field \cite{Normand1992}. Additionally, in thermally activated transport near the threshold voltage, the Lorentz force will become comparable to the electric field and hence, the so-called gyrotropic tunneling of quasiparticles \cite{Laitinen2018} will influence the sliding of the Wigner crystal. For example, by using a $45^{\circ}$ sliding angle \textit {w.r.t.} to radial direction, a perfect fit using $N = 3$, $T = 0.4$ K is achieved for our experimental non-linear \textit {IV} data, which removes the discrepancy between the microwave absorption and \textit {IV} measurements. Nevertheless, the average domain size $L_c >$ 3 {\it a}, where {\it a $\sim 1/\sqrt{\pi n}$} is the Wigner crystal lattice constant, verifies that well-defined local crystalline order persists in our sample.

One factor enhancing the observability of the Wigner crystal in suspended graphene may be provided by the deformation of the graphene sheet into a dimple lattice under the action of the back gate voltage (see the inset of Fig. 4C). This would add inertia to the charge carriers, as is the case with electrons on superfluid helium where the Wigner crystal has been observed even in the absence of magnetic field \cite{Shirahama1995}. Frozen-in dimples, on the other hand, may act as important pinning sites, in addition to the charged pinning centers provided by residual charge density (see supplemental information).

Our experiment clearly supports the presence of crystalline Wigner solid order in region 1/7 $< \nu <$ 1/5, whereas a strongly gapped state is observed at $ \nu \ll 1/7$. This gapped state around $\nu$ = 0 may be related to excitonic ordering. The crossover between these two regimes appears to take place at $\nu \simeq 0.14$, around which we find activated behavior with a gap of $\Delta_{\nu = 0.14} \simeq$ 1 K (see Fig. 3B). Re-entrant behavior of the FQH order is one possibility for the observed behavior, but our experiment does not resolve any clear characteristics associated with a particular fractional state.

\section*{Discussion}
In suspended graphene, impurities and the back-gate-modified dimple lattice create disorder potentials which are screened by charge, leading to a constant electrochemical potential. But at very low charge density, the effectiveness of the charge screening breaks down, exposing hills and valleys of the disorder potential. The electrons will be localized in the valleys of the potential landscape, and they become rearranged in reduced dimension via Coulomb interactions. Owing to the reduced screening and dimension, the charges will interact strongly. In an intermediate regime where the kinetic energy is still appreciable, the composite fermions will condense to fractional states such as $\nu = 4/13$ and $\nu = 4/11$ in partially filled Landau levels. But at even lower charge density, the Coulomb interaction energy will strongly dominate over the kinetic energy, leading to freezing of the electronic states into more ordered states such as the Wigner crystal pinned by the potential landscape.  Hence, the formation of exotic FQHE and WC states do both owe to the reduced dimensionality and higher Coulomb interactions in suspended graphene where, besides, the lattice of dimples could be a significant contributing factor. The experiment demonstrated here opens up the possibility for probing new kind of strongly correlated electron states which may emerge more markedly in suspended graphene than in any other system.

In conclusion, we have investigated fractional quantum Hall ordering in two-terminal suspended graphene Corbino devices where no edge states carry current across the sample. For bulk transport in the Corbino geometry, our measurements present an unparalleled sequence of FQH states in transport measurements on two-lead graphene. Besides many regular FQH states, we also resolved unconventional fractional quantum Hall states at filling factors $\nu$ = 4/13 and $\nu$ = 4/11. The presence of these states indicates residual interaction between composite fermions. At low filling factors we observe competition between liquid and solid ordering. Our results, most notably microwave absorbtion spectra, support Wigner crystallization of electrons in region 1/7 $< \nu <$ 1/5 with a melting temperature of 1.5 K. The cross-over from the strongly-gapped quantum Hall state at $\nu = 0$ to the Wigner crystal regime, is found to involve an intermediate state with reduced incompressibility. Being a truly 2D system, graphene provides new possibilities for studies of ordered electron gas/liquid/solid phases with the added benefit that the structures exist right at the surface, fully exposed for probing.

\section*{Acknowledgements}
We thank C. Flindt, A. Harju, Y. Meir, T. Ojanen, S. Paraoanu, E. Sonin, and G. F\'eve for fruitful discussions. This work has been supported in part by the EU Framework Programme (FP7 and H2020 Graphene Flagship), by ERC (grant no. 670743), and by the Academy of Finland (projects no. 250280 LTQ CoE and 286098). A.L. is grateful to Vais{\"a}l{\"a} Foundation of the Finnish Academy of Science and Letters for a scholarship. \textbf{Author contributions:} The research was initiated by P.J.H. The experimental principle was selected by M.K. and P.J.H. The sample structure was developed and manufactured by A.L. The experiments were carried out by A.L. and M.K who were also responsible for the data analysis.  The results and their interpretation were discussed among all the authors. The paper and its supplement were written by the authors together. \textbf{Competing interests:} The authors declare that they have no competing interests. \textbf {Data and materials availability:} All data needed to evaluate the conclusions in the paper are present in the paper and or the supplemental information. Additional data related to this work available from authors upon request.

\section*{Materials and Methods}
\subsection*{Sample fabrication}

Our sample fabrication is based on well-chosen combination of resists with differential selectivity, which allowed the fabrication steps for deposition of a suspended top contact (see supplemental material for details ). We exfoliated graphene (Graphenium form NGS Naturgraphit GmbH) using a heat-assisted exfoliation technique to maximize the size of the exfoliated flakes~\cite{Huang2015}. Monolayer graphene flakes were located by their contrast in an optical microscope and verified using a Raman spectrometer with He-Ne laser (633 nm). The first contact defining the outer rim of the Corbino disk (see Fig. 1)  was deposited in the usual manner \cite{Tombros2011}, and later the inner contact was fabricated along with an air bridge to connect the inner contact to a bonding pad (Fig. S1). Strongly doped silicon Si++ substrate with 285 nm of thermally grown SiO$_2$  was used as a global back gate.

Due to the  exposure of graphene to resists during the fabrication process, our suspended devices tend to be highly doped (mostly p-type) in our first resistance $R_d = dV/dI$ \textit{vs.} gate voltage $V_g$ scans.
Annealing of samples on LOR was typically performed at a bias voltage of 1.6$\pm$0.1 V which is quite comparable with our HF etched, rectangular two-lead samples \cite{Laitinen2014}. For further details on annealing, we refer to the supplemental information (Fig. S2a).

\subsection*{Measurement techniques}
Our measurements down to 20 mK were performed in a BlueFors LD-400 dilution refrigerator. The measurement lines were twisted pair phosphor-bronze wires supplemented by three stage $RC$ filters with a nominal cut-off given by $R=100$ Ohms and $C=5$ nF. However, due high impedance of the quantum Hall samples the actual cutoff is determined by the sample resistance.  For magnetoconductance measurement, we used an AC peak-to-peak current excitation of $0.1$ nA at $f=3.333$ Hz.

For transconductance  $g_m=dI/dV_g$ we measured both magnitude $\textrm{Mag}\{g_m\}$ and phase $\textrm{Arg}\{g_m\} $ using low frequency lockin detection.  The best results for $g_m$ correlation analysis were obtained by recording $\textrm{Arg}\{g_m\} $ at a bias voltage $V$ that corresponded to the onset of the $V^{\alpha}$ regime, see supplemental information Fig. S7a. Consistent information was obtained by analyzing the simultaneously acquired $\textrm{Mag}\{g_m\}$. The gate frequency in AC transconductance measurements was set at $f$ = 17.777 Hz, while the peak-to-peak  AC excitation amplitude was adjusted to correspond to charge of one electron over the sample. The DC bias between source and drain was varied in the range $V$ = 0.1 - 0.5 mV. Note that this technique was mainly applied to the Jain sequence states as the unconventional states ($\nu = 4/13$ and $\nu = 4/11$) were easily overshadowed by the close dominant $\nu=1/3$ state in the cross-correlation analysis due to the nature of the technique, for further information see the supplemental information.


\clearpage
\clearpage
\begin{table}[h!]
\caption{Characteristics of our suspended Corbino graphene samples (by columns): inner and outer radii $r_i$ and $r_o$, respectively, $n_0$ is the residual charge density,  field effect mobility $\mu_f$ is measured at $n \sim 5 \times 10^{10}~\textrm{cm}^{-2}$, and $\sigma_{0}$ denotes the minimum conductivity; the contact resistance $R_C$ was subtracted off from the conductivity only when calculating $\mu_f$. The FQH states probed by $g_m$ are indicated by square brackets [p/q], otherwise the identification is based on differential magnetoconductance $G_d=dI/dV|_{V=0}$. The label \text{EV$^*$} corresponds to a set of measurements done on sample EV in its second cool down, in which a change in FQH state quality is observed along with decreased mobility.}
 \label{Tab1}
\centering
\begin{tabularx}{\textwidth}{XX}
\begin{tabular}{c c c c c c c}
\\
\hline
\rowcolor{white}
Sample &$r_i/r_o$&$n_o$ & $\mu_f$ & $\sigma_0 $ & Measurement & FQH states \\
\rowcolor{white}
 &in $\mu$m& $1/\textrm{cm}^2$ & $\textrm{cm}^2/\textrm{Vs}$ & $\textrm{S}$ &  & \\
\hline \hline
\hline
\\

\mbox{\textbf{XD}}& \mbox{0.35/1.0} &\mbox{$2 \times 10^{10}$} & \mbox{$4.0\times 10^4$} & \mbox{$1.6 \times 10^{-5}$} & \mbox{$g_m$ and $G_d$} & \mbox{$\{ \frac{1}{3}, [\frac{2}{5}], [\frac{4}{7}], \frac{2}{3}, [\frac{4}{5}], [\frac{4}{3}]\}$}\\
\mbox{\textbf{EV}}& \mbox{0.40/1.6}&\mbox{$6 \times 10^{9}$} & \mbox{$1.2 \times 10^5$} & \mbox{$1.4 \times 10^{-4}$} & \mbox{$g_m$ and $G_d$} & \mbox{$\{[\frac{2}{7}], \frac{1}{3}, \frac{4}{11}, \frac{2}{5}, \frac{3}{7}, [\frac{4}{9}], \frac{4}{7}, \frac{3}{5}, \frac{2}{3}\}$}\\
\mbox{\textbf{EV$^*$}}& \mbox{0.40/1.6}&\mbox{$6 \times 10^{9}$} & \mbox{$1.1 \times 10^5$} & \mbox{$1.1 \times 10^{-4}$} & \mbox{$G_d$ and $\delta G_d/\delta n$} & \mbox{$\{\frac{1}{3}, \frac{2}{5}, \frac{3}{7}, \frac{3}{5}, \frac{2}{3}, \frac{4}{3}\}$}\\
\mbox{\textbf{EV$_2$}}& \mbox{0.75/1.9}&\mbox{$5 \times 10^{9}$} & \mbox{$1.3 \times 10^5$} & \mbox{$2.0 \times 10^{-4}$} & \mbox{$G_d$ and $\delta G_d/\delta n$} & \mbox{$\{\frac{1}{5}, \frac{2}{7}, \frac{4}{13}, \frac{1}{3}, \frac{4}{11}, \frac{2}{5}, \frac{3}{7}, \frac{4}{7}, \frac{3}{5}, \frac{2}{3}, \frac{4}{3}\}$}\
\\
\\
\hline
\end{tabular}
\end{tabularx}
\end{table}

\clearpage
\noindent {\bf Fig. 1.} Suspended graphene Corbino disk: Scanning electron micrograph of a 2-$\mu$m-diameter graphene Corbino sample (XD) with 700 nm diameter circular middle contact; blue denotes the suspended graphene and gold/chromium contacts appear as grayish yellow. See the supplemental information for detailed structure (Fig. S1).

\noindent {\bf Fig. 2.} Transconductance measurement: (A) The derivative of logarithmic differential magnetoconductance {\it w.r.t.} $n$, {\it i.e.} $\delta \log (G_d)/\delta n$, of sample EV$^*$ (see Table I), displayed as a Landau fan diagram on the $B$ vs. charge density $n$ plane. The blue  lines at $\nu = nh/eB$ identifying regular QH states $\nu$ = \{2, 6, 10\}, broken symmetry states $\nu$ = \{0, 1, 3\} originating from lifting spin and valley degeneracies, and FQH states $\nu = $\{-1/3, 1/3, 2/5, 3/7, 3/5, 2/3, 4/3\}. (B) Modulation of transconductance phase $\textrm{Arg} \{g_m\}$ measured for sample EV over charge density $n$ ($V_g$ up to 15 V) at magnetic fields between 6 and 8 T (filling factor $\nu = 1$ is reached at the lowest right corner). The data are divided into ten regions, each marked with a color bar at the top of the stripe: we use eight vertical stripes of equal size and two inclined stripes which are defined in order to improve the contrast of fringes at $\nu$ = 3/7 and $\nu$ = 2/3. The correlation analysis has been performed separately in each region. The correlation function $h(\nu)$ specifies the structure of these fringes {\it w.r.t.} the slope $\delta n/\delta B$ equaling to the filling factor $\nu$. The black dashed lines off the borders indicate FQH states with the marked fractional filling factor. (C) The scaled cross-correlation function $c(\nu) = h(\nu)/h(\nu)_{max}$, for all values of $\nu = 0.2 \ldots 1.1$, identifying fractional fillings $\{2/7, 1/3, 2/5, 3/7, 4/9, 4/7, 3/5, 2/3\}$, in agreement with our magnetoconductance measurements; see the supplemental information for analysis (Fig. S4). The color of the distribution refers to the color code of the analyzed regions in Fig. 2B. The peak values are marked with the corresponding fractional filling factors. Further FQH states are displayed in inverse differential conductance data in Fig. 3 (see also supplemental information Fig. S5). The measurement was performed at mixing chamber temperature of 20 mK.

\noindent{\bf Fig. 3.}  Unconventional FQH states: (A) Inverse differential conductance through the Corbino disk measured at AC for sample EV2 at magnetic fields between 6 - 8.5 T, measured at 20 mK. The magnetic field varies from  8.5 T (blue) to 6 T(red). The fractional states are identified by peaks at $\nu = nh/eB$ in inverse conductance traces which collapses on top of each other in $G^{-1}_d$ vs $\nu$ plot. Distinct states at $\nu$ = $\{4/13, 1/3, 4/11, 2/5\}$ are identified from measured traces. (B) The Arrhenius plot of DC conductivity $\sigma_{xx}$ at $B$ = 7 T for fractional states at $\nu$ = 4/11 ($\lozenge$) and $\nu$ = 4/13 (\textcolor{red}{$\ocircle$}), contrasted to an activation measurement at $\nu$ = 0.14 (\textcolor{green}{$\triangledown$}) at $B$ = 9 T.  The straight lines represent Arrhenius fits that were used to extract the gaps of these states. The obtained gaps are also displayed as an inset.

\noindent{\bf Fig. 4.} Wigner crystal: (A) The absorption microwave spectra of the detected absolute value of the dc current $I$ vs irradiation frequency $f$ for sample EV2, measured at $\nu$ = 1/3 (red), $\nu$ = 0.16 (green), $\nu$ = 0.15 (blue) and $\nu$ = 0.12 (yellow),  and  at magnetic field {\it B} = 9 T and rf-power = -68 dBm. An absorption resonance  with a maximum response for $\nu = 0.16$, is seen at frequency 3.0 GHz which is identified as a collective mode frequency in the disordered-pinned Wigner crystal phase (see text). The flat curve at $\nu$ = 0.12 is an indication of a strongly gapped fractional quantum Hall state. (B) Microwave absorption spectra in terms of dc current $I$ measured across the $\nu = 0.16$, (at fixed charge density, $V_g$ = 2.8 V, rf-power = -65 dBm, {\it B} =\{9 T (blue), 8.8 T (black), 8.6 T (yellow), 8.4 T (violet), 8 T (red)\}). Broadening of the resonance peak is seen clearly at $B=8$ T, because the WC is closest to its melting temperature at this field. Note the increase in the dc current compared to Fig. 4A  (see also Fig. S10 in the supplemental information) (C) Semilog plot of two IV curves from sample XD corresponding to filling factors in frame (a): (\textcolor{red}{$\circ$}) $\nu = 0.16$ and (\textcolor{blue}{$\square$}) $\nu$ = 0.33. The black curve denotes a fit of $I_W$ (see Eq. \ref{eq:depinning}) describing electrical transport in the Wigner crystal regime by thermally activated depinning of crystallites, while the orange curve displays linear IV behavior with a total resistance of 8 M$\Omega$ due to quantum tunneling over a chain of localized states. This is clearly different from Wigner crystal behaviour and well valid up to 1 mV. $V_T$ marks the threshold voltage. The inset displays a disordered Wigner crystal where charge ordering follows corrugations of suspended graphene: higher charge density (yellow) resides at the dimples. (D) The temperature dependence of the microwave absorption peak of Fig. 4A at the Wigner crystal filling factor $\nu$ = 0.15. The maximum in the current indicates the melting point of the Wigner crystal order.

\begin{figure}[ht]
\centering
\includegraphics[width=1\linewidth, center]{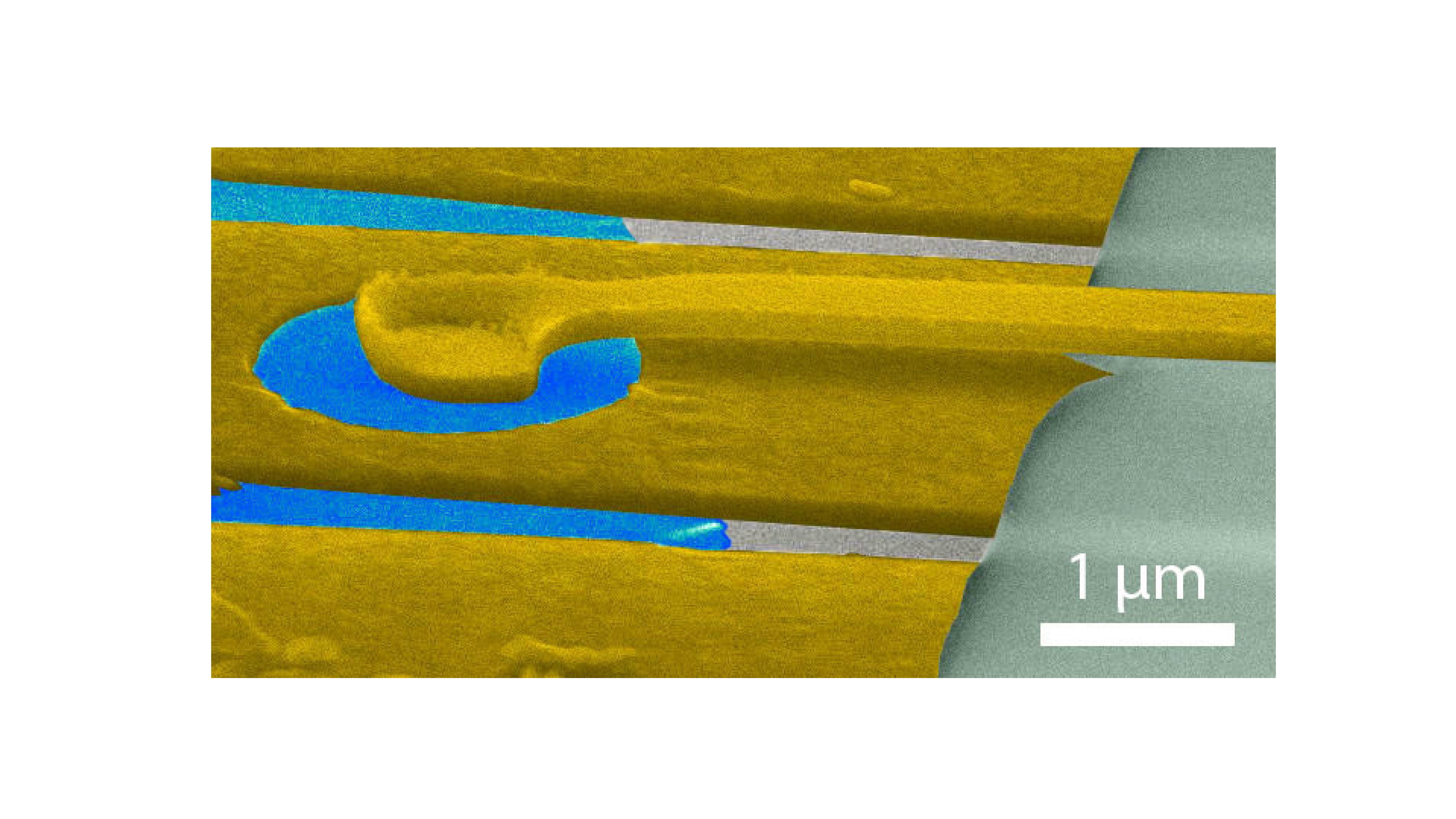}
\caption{Suspended graphene Corbino disk}
\end{figure}

\begin{figure}[ht]
\centering
\includegraphics[width=1\linewidth, center]{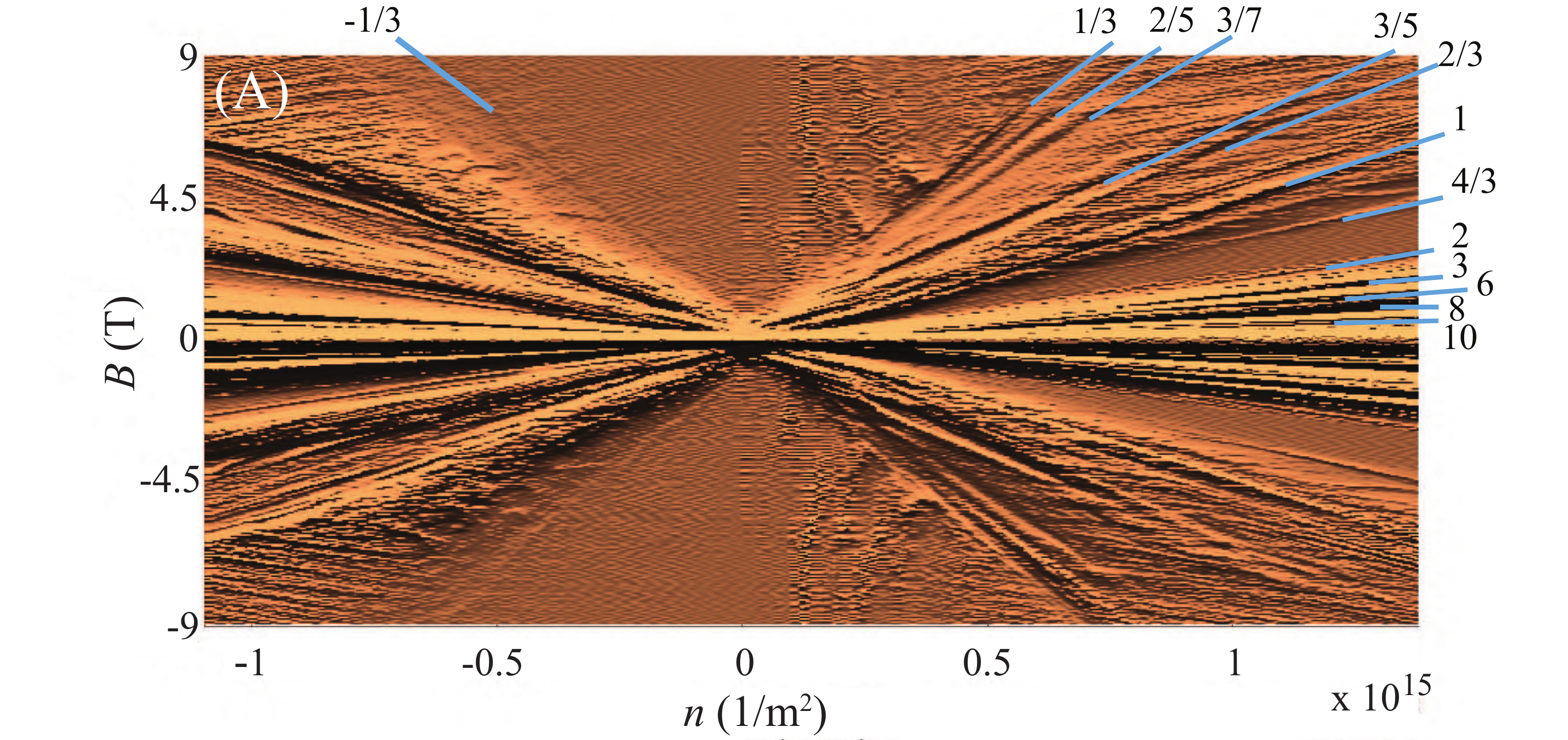}

\centering
\begin{minipage}{0.52\linewidth}
\includegraphics[width=1\linewidth, center]{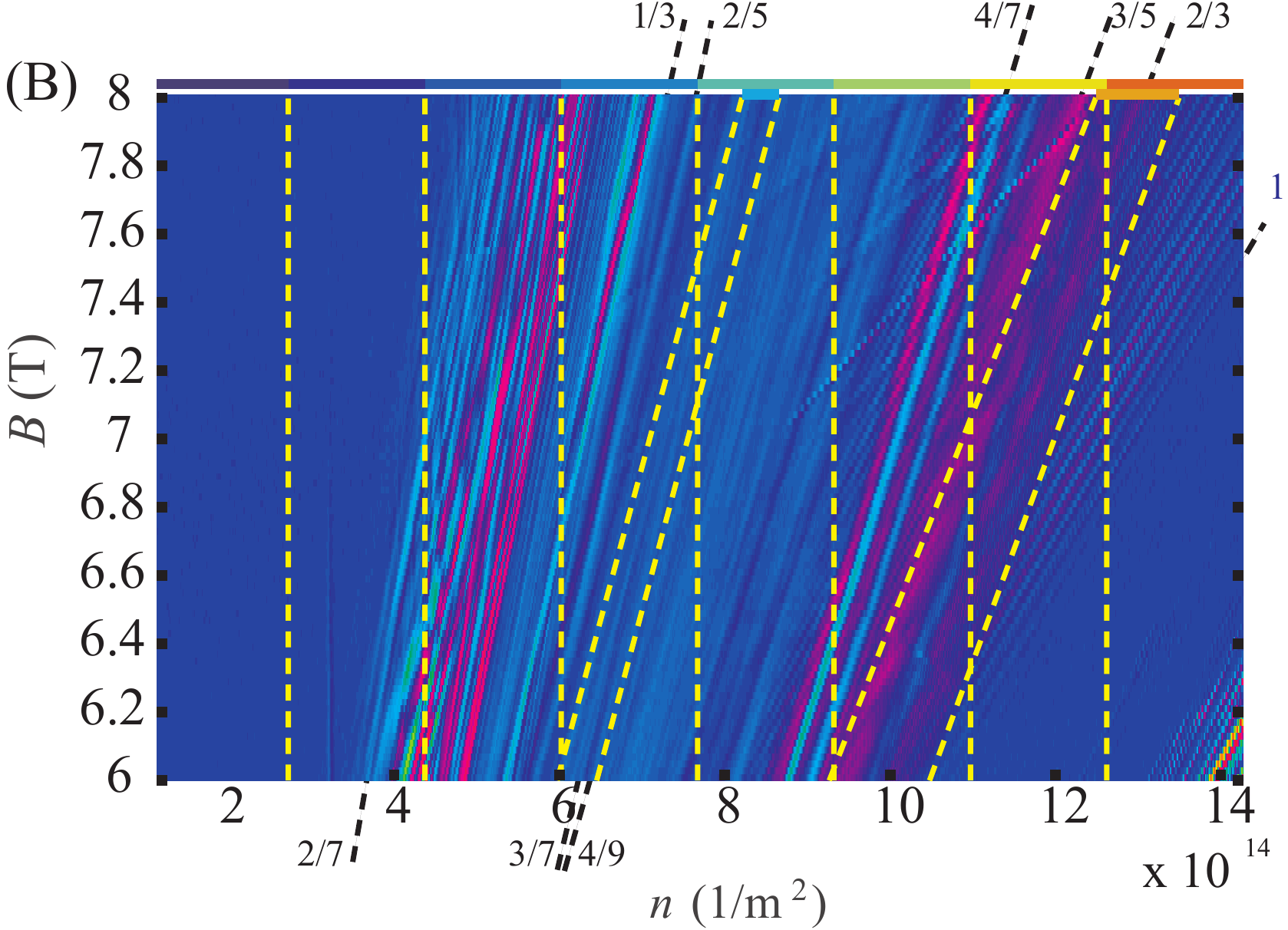}
\end{minipage}
\hfill
\begin{minipage}{0.47\linewidth}
\centering
\includegraphics[width=1\linewidth, center]{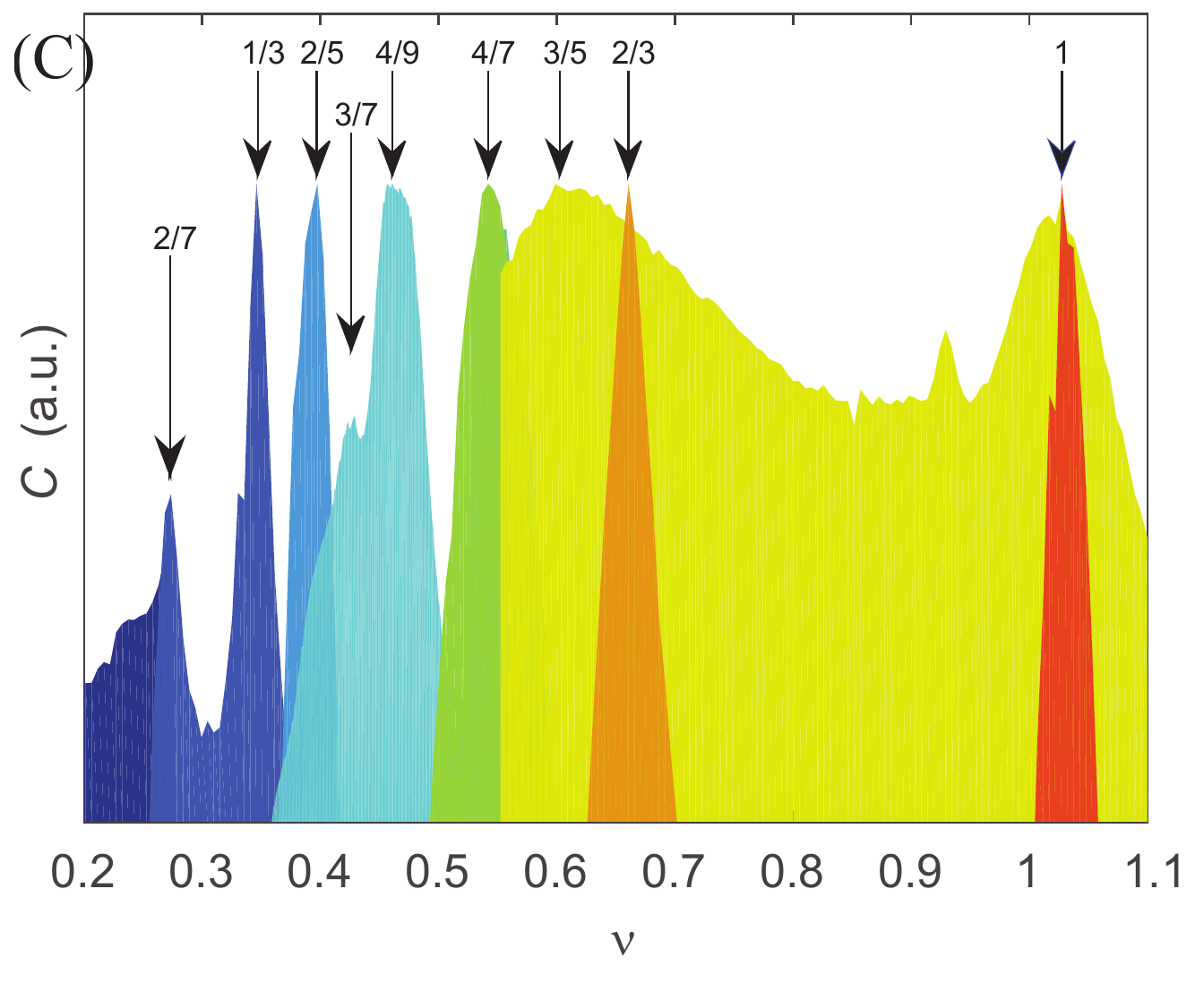}
\end{minipage}
\caption{Transconductance measurement}
\end{figure}

\begin{figure}[ht]
\centering
\begin{minipage}{0.4\linewidth}
\includegraphics[width=1\linewidth, center]{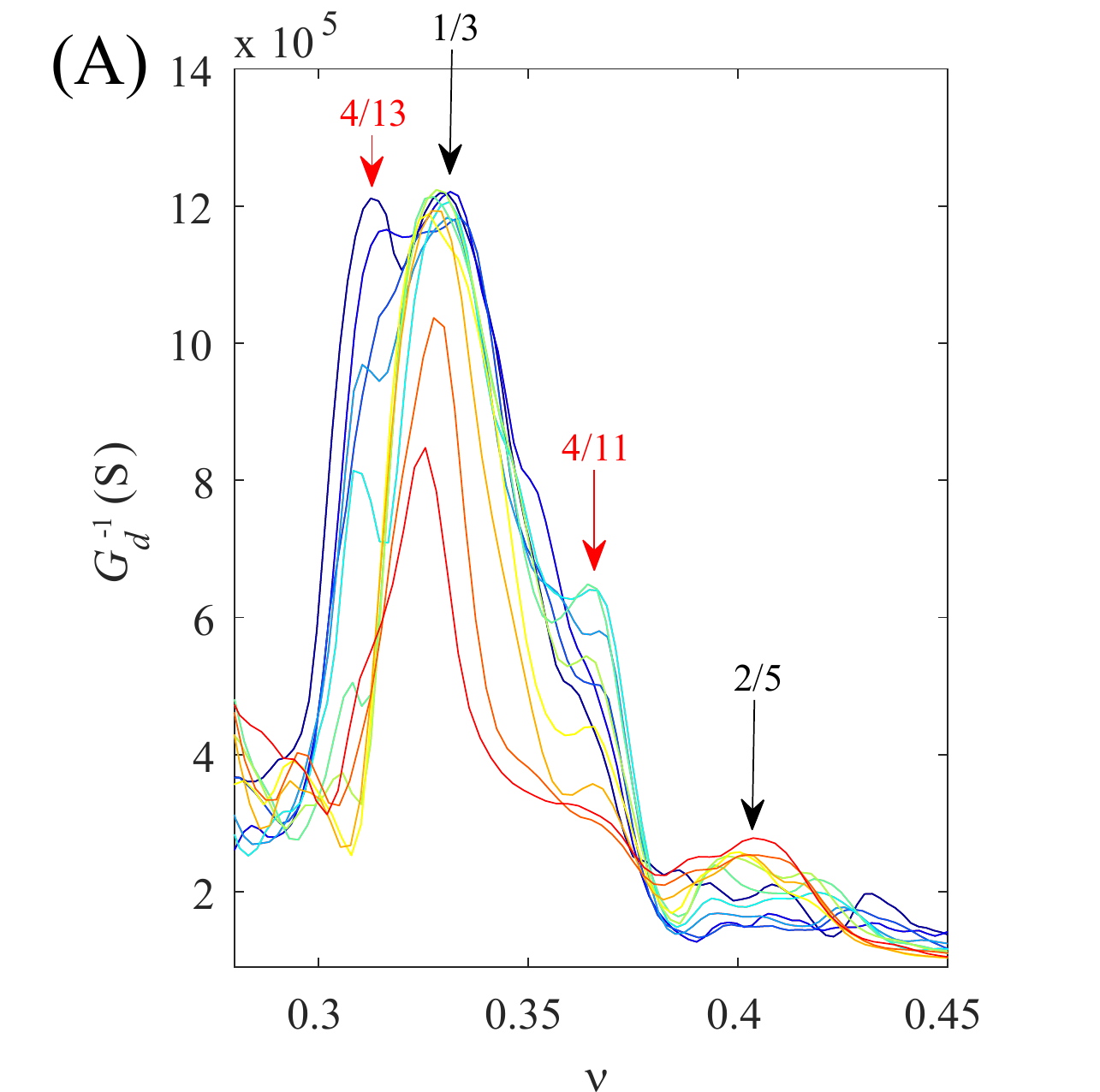}
\end{minipage}
\begin{minipage}{0.5\linewidth}
\centering
\includegraphics[width=1\linewidth, center]{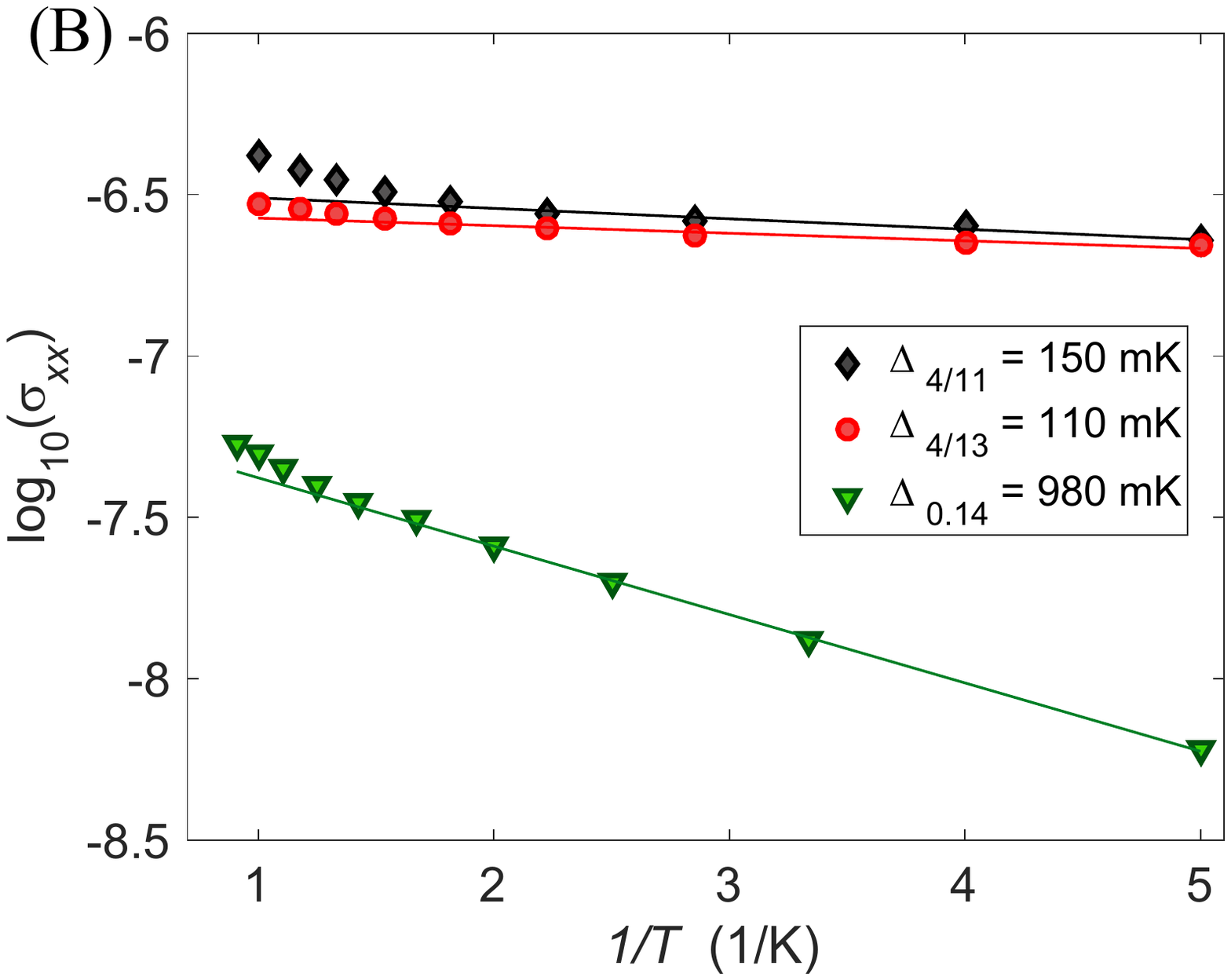}
\end{minipage}
\caption{Unconventional FQH states}
\end{figure}

\begin{figure}[ht]
\centering
\begin{minipage}{0.45\linewidth}
\includegraphics[width=1\linewidth, center]{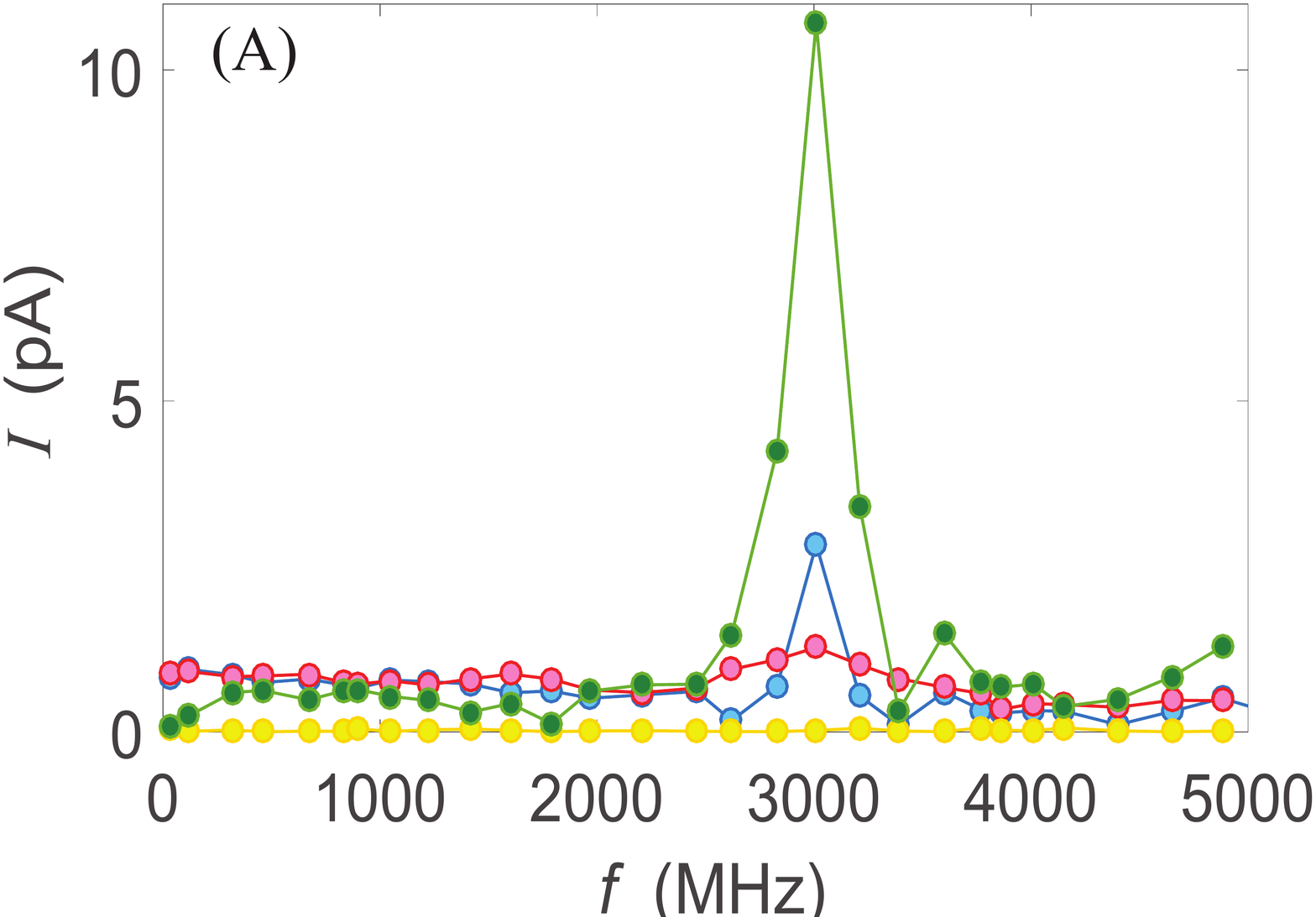}
\end{minipage}
\hfill
\centering
\begin{minipage}{0.50\linewidth}
\includegraphics[width=1\linewidth, center]{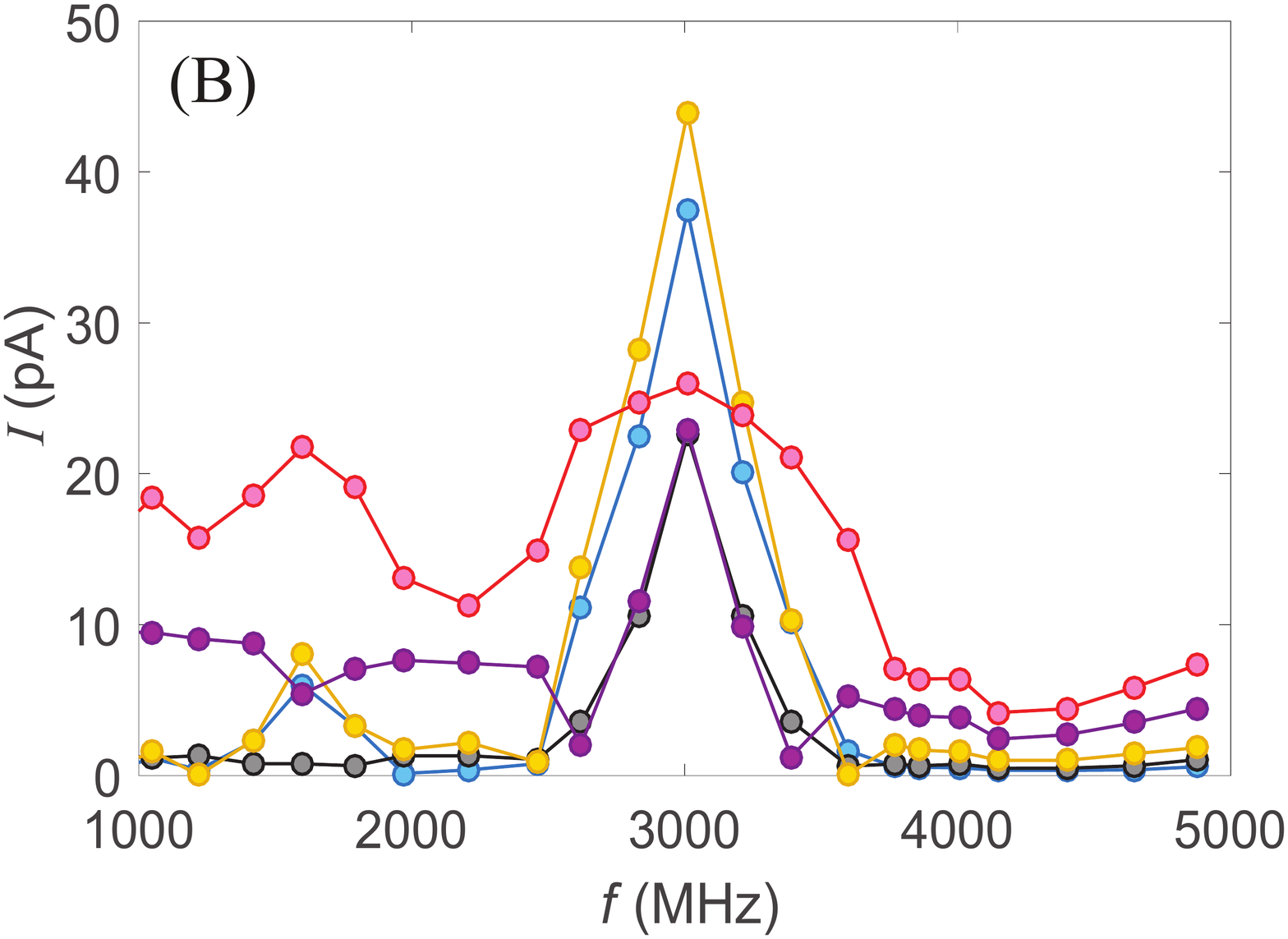}
\end{minipage}
\hfill
\centering
\begin{minipage}{0.50\linewidth}
\centering
\includegraphics[width=1\linewidth, center]{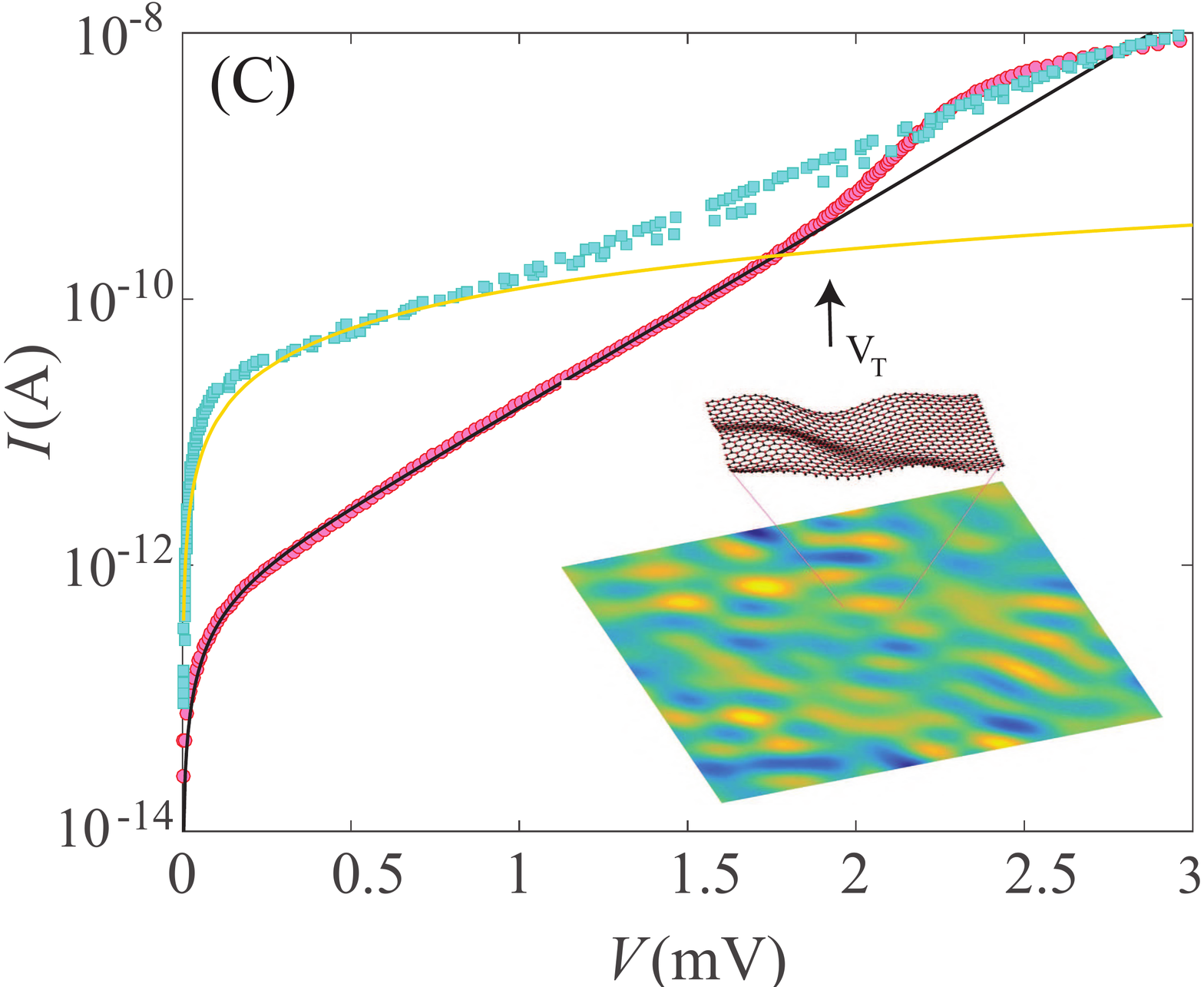}
\end{minipage}
\centering
\includegraphics[width=0.45\linewidth, center]{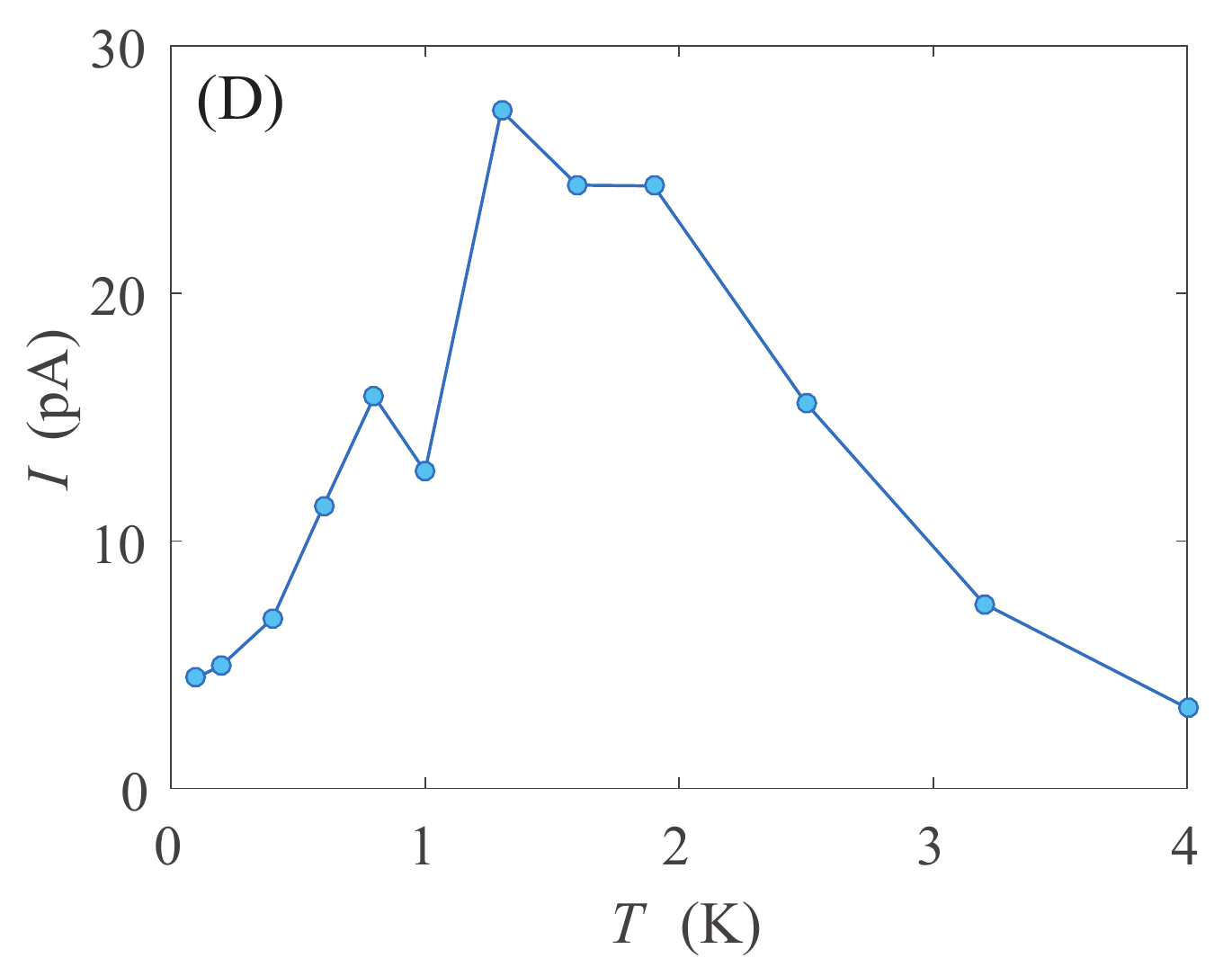}
\caption{Wigner crystal}
\end{figure}

\clearpage 



\section*{Supplementary material}

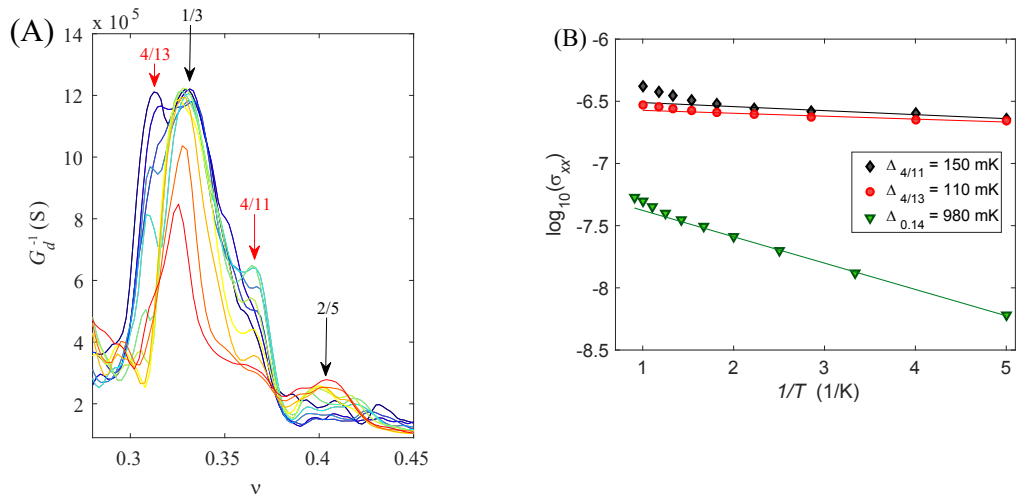

FIG. 3. Unconventional FQH states

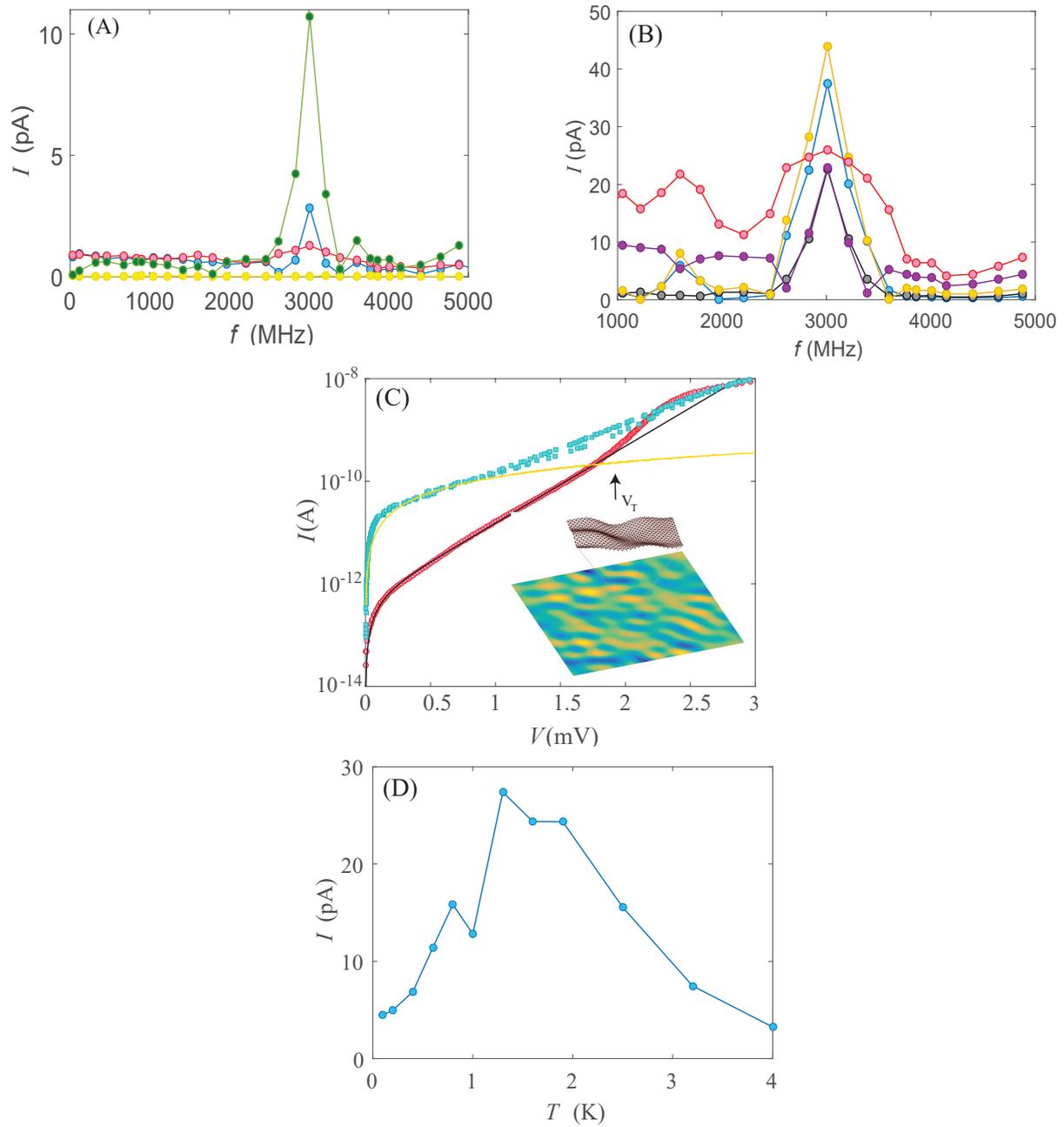

FIG. 4. Wigner crystal

23

In this supplement we provide additional information on sample fabrication, current annealing of graphene to enhance mobility, sample characterization, the identification of fractional states using transconductance, transport mechanisms in our samples, rf-measurements of Wigner crystal, and the nature of its pinning.

## S1    Sample Fabrication

Our sample fabrication is based on well-chosen combination of resists with differential selectivity, which allowed the fabrication steps for the deposition of a suspended top contact. In this work, we employed heat assisted mechanical exfoliation technique (*1*). The application of heat assisted exfoliation on top of LOR resist produced large graphene flakes, which was the basic requirement for a good Corbino device. A $250 + 250$ nm thick base layer of LOR-3A (*2*) was spun in two steps on $SiO_2$/Si++ and baked at $200°$C for 5 and 12 min, respectively. A graphite covered tape was pressed against the LOR-covered wafer for 1 min and subsequently baked on a hot plate at $100°$C for 2 min. The tape was peeled from the chip after it had cooled down to room temperature. The exfoliated graphene flake was first characterized optically and then using Raman spectroscopy at 633 nm.

In the first lithography step, the lead for the outer ring and its contact pad were fabricated. A double layer PMMA (950K/A3: top layer and 50K/A11: bottom layer) spun and patterned using e-beam. A Cr/Au sandwich of $5/70$ nm thickness was evaporated at pressure $p < 6 \times 10^{-8}$ mBar. After metal deposition, special care was taken during the metal lift-off process: the lift off was performed in $80°$C xylene and the unwanted residual metal from the circular hole of outer lead was removed using micromanipulation. The outer ring electrode is of rectangular geometry extending over the graphene flake with a circular hole of diameter $2 - 10$ $\mu$m, see Fig. S1c. On the processed sample, another 500 nm thick LOR was spun. A via in the center of the circular hole of outer lead was opened in the top LOR layer using e-beam lithography. A dose of $1000$ $\mu$C/cm$^2$ was used for this step. The exposed pattern was developed in ethyl lactate for 60 s and rinsed in hexane. The diameter of the via varied between $0.5 - 5$ $\mu$m, see Fig. S1d. This via acts as foundation for the inner electrode contact in the concentric Corbino geometry. For fabrication of the inner lead, a double layer PMMA of 200 nm was spun and patterned using e-beam lithography. On the patterned device, Cr (5 nm)/Au (100 nm) was deposited, and the lift off was done in hot xylene with a rinse in hexane, see Fig. S1e.

In final steps of the lithography, the graphene between inner and outer ring leads was suspended by exposing LOR to a dose of $1000$ $\mu$C/cm$^2$. In order to

expose the outer lead bonding pad, LOR on the larger structures was exposed to an e-beam dose of 700 $\mu$C/cm$^2$. The device was developed and cleaned in ethyl lactate for 60s and rinsed in hexane for 60s. A schematic cross section of the final device is depicted in Fig. S1f and an SEM image of a complete sample structure is shown in Fig. 1 of main text. For bonding, part of the pad was pressed against the SiO$_2$, as shown in Fig. S1g. These pressed pads were used for bonding. The finished devices were electrically characterized at room temperature, and devices with low resistance were selected for low temperature measurements.

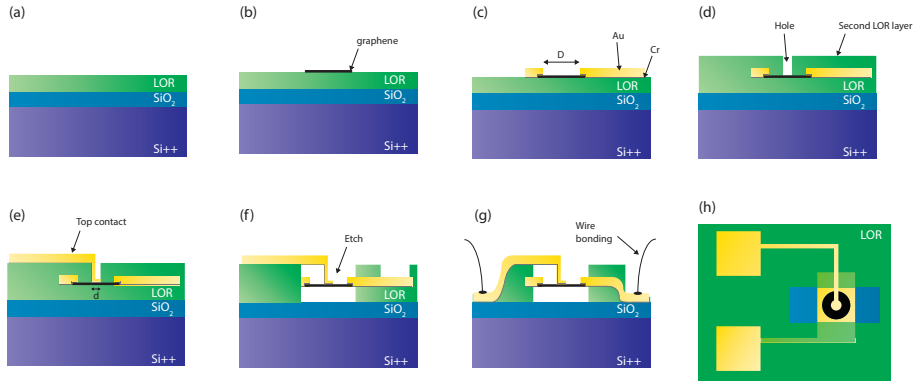

Figure S1: Fabrication process steps: a) LOR layer with thickness of ∼500 nm is spun on a cleaned Si/SiO$_2$ chip. b) Mechanical exfoliation of graphene on top of the LOR layer. c) Fabrication of outer lead and its bonding pad. d) Formation of a via for the inner lead using a second LOR layer. e) Air bridge connecting inner lead and pad is patterned using an e-beam lithography step with PMMA. f) Suspension of graphene between the inner and outer leads and uncovering of the outer lead-pad by e-beam exposure. g) Status after wire bonding: gold pads are deformed and pushed partly down to silicon. h) A schematic top view of a sample with suspended top contact: black color shows suspended graphene, blue color shows the exposed window in LOR, light yellow color shows the covered outer lead and bright yellow color indicates clean gold of the inner lead and the outer leads.

## S2   Sample characterization

Our measurements down to 20 mK were performed in a BlueFors LD-400 dilution refrigerator. The measurement lines were twisted pair phosphor-bronze wires supplemented by three stage $RC$ filters with a nominal cut-off given by $R = 100$ Ohms

and $C = 5$ nF. However, due high impedance of the quantum Hall samples the actual cutoff is determined by the sample resistance. For magnetoconductance measurement, we used an AC peak-to-peak current excitation of 0.1 nA at $f = 3.333$ Hz.

For transconductance $g_m = dI/dV_g$ we measured both magnitude Mag$\{g_m\}$ and phase Arg$\{g_m\}$ using low frequency lock-in detection. The best results for $g_m$ correlation analysis were obtained by recording Arg$\{g_m\}$ at a bias voltage $V$ that corresponded to the onset of the $V^\alpha$ regime. Consistent information was obtained by analyzing the simultaneously acquired Mag$\{g_m\}$. The gate frequency in AC transconductance measurements was set at $f = 17.777$ Hz, while the peak-to-peak AC excitation amplitude was adjusted to correspond to charge of one electron over the sample. The DC bias between source and drain was varied in the range $V = 0.1$ - 0.5 mV.

## S2.1   Annealing

Our suspended Corbino graphene samples were predominantly of $p$-type. During typical annealing process, the doping shifts to $n$-type before reaching the desired charge-neutral state. For achieving charge neutrality, a high current annealing was performed. We employed voltage bias for our current annealing process: In contrast to current-biased devices, voltage-biased samples are protected against possible rise in resistance due to shifting of the Dirac point. A typical bias voltage value used in annealing of our samples was $1.6 \pm 0.1$ V; in spite of strong heating of the refrigerator, the sample was kept at temperatures $T < 1$ K during annealing. Slight asymmetry in the concentricity of the electrodes in our Corbino geometry, inverse radial decay of current, and local variation of the contact resistance result in non-uniformity of current annealing. Owing to this non-uniformity, our suspended Corbino disk becomes split into differently annealed graphene domains, each having their own charge neutrality point. This can be seen in Fig. S2a. Eventually, during annealing process several of these peaks merge to give a single Dirac peak or closely degenerate Dirac peaks. See Fig. S2b for the final $R_d$ vs $V_g$ curve of the sample EV* after annealing.

## S2.2   Residual Doping

The level of residual charge doping was extracted from $G$ vs $n$ -plots on *loglog*-scale as the point where conductance $G$ levels off when approaching the Dirac point. We employed straight line fits (power laws with exponent around 1/2) to the conductance data in order to better define the position of this cross-over point, as shown in Fig. S2c.

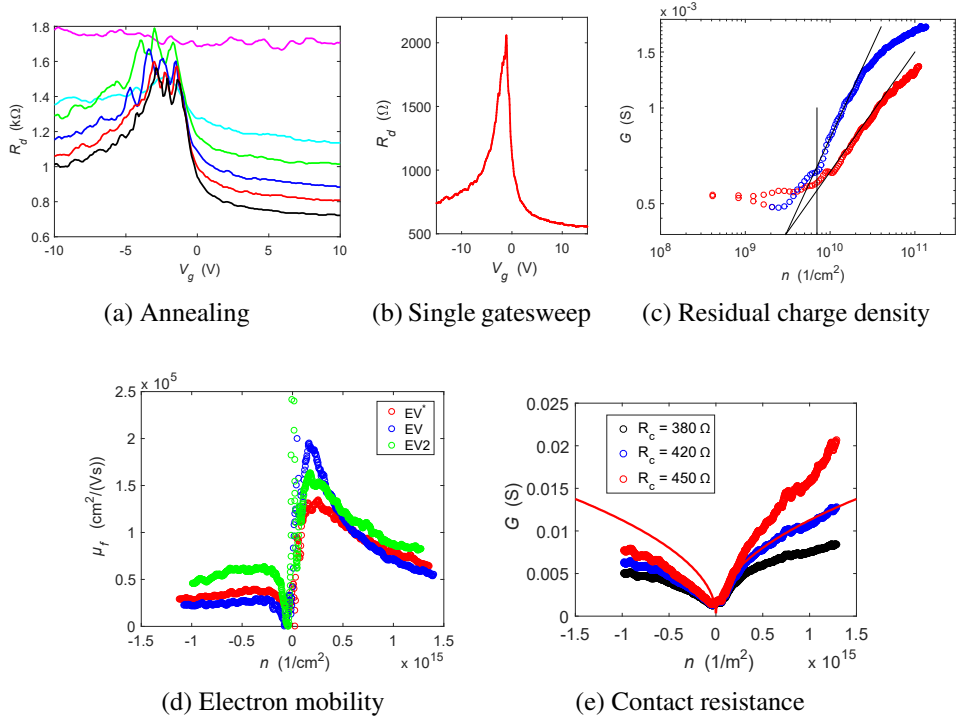

(a) Annealing    (b) Single gatesweep    (c) Residual charge density

(d) Electron mobility    (e) Contact resistance

Figure S2: Sample characterization : a) Annealing progress of a Corbino sample with outer diameter $D = 3.2$ μm and inner contact diameter $d = 0.8$ μm. The magenta curve at the top is the first differential resistance $R_d$ trace measured after annealing current $I \approx 1.5$ mA ($V \approx 1.3$ V) while the black curve, obtained after $I \approx 2$ mA ($V \approx 1.7$ V), denotes the final annealing result (sample EV in Table I of the main paper). b) $R_d$ vs $V_g$ curve for sample EV* after annealing. c) $G$ vs $n$ for sample EV* on *loglog*-scale: red and blue circles represent data on the hole and electron sides, respectively. The black lines represent power law trends in the data. The cross-over point of the curves from power law to constant conductance is taken as $n_0$, which yields here $n_0 = 6 \times 10^9$ cm$^{-2}$. $R_c$ was not subtracted in this analysis. d) Mobility as a function of electron density $n$ ($n < 0$ corresponds to holes) determined for samples EV*, EV, and EV2 at zero magnetic field. Contact resistance $R_c$ (see text) has been subtracted off from the measured resistance, *i.e.*, $G = \frac{1}{R_d - R_c}$ in this analysis. e) DC conductance measured for sample EV2 assuming an external series resistance, a contact resistance of $R_c = 380$ Ω (black circles), $R_c = 420$ Ω (blue circles), and $R_c = 450$ Ω (red circles). The solid red curve illustrates a theoretically calculated curve which agrees with the data when the contact resistance is set to $R_c = 420$ Ω, *i.e.*, $G = \frac{1}{R_d - R_c}$ for the data points.

4

The residual doping for other samples was extracted in the same way: $n_{0,\text{XD}}$ = $2 \times 10^{10}$ cm$^{-2}$ and $n_{0,\text{EV}}$ = $6 \times 10^9$ cm$^{-2}$. The residual density modulation $n_0$ amounts to about 15 and 30% of the Wigner crystal charge density $n = 2.6 - 4.4 \cdot 10^{10}$ cm$^{-2}$ in samples EV and XD, respectively.

## S3   Mobility and Contact resistance

Field effect mobility of our samples was calculated using the formula

$$\mu_f = \frac{\sigma - \sigma_0}{ne},$$   (S1)

where $\sigma_0$ is the measured minimum conductivity, and $n$ denotes the charge carrier density. The conductivity in Corbino geometry can be calculated from measured conductance $G$ using

$$\sigma = G \frac{\ln(r_o/r_i)}{2\pi},$$   (S2)

where $r_o$ and $r_i$ are the radii of outer and inner contacts, respectively.

When calculating the mobility using Eqs. S1 and S2, we take into account the contact resistance $R_c$ . We estimate $R_c$ from the comparison of measured data to the theory based on evanescent mode transport (3). Fig. S2e (see Sect. 4.1) displays such a fitting for sample EV2, which yields $R_c$ = 420 $\Omega$ for the contact resistance. Similar comparison was done for EV and EV*, and we obtained $R_c$ = 410 $\Omega$ for EV* and $R_c$ = 550 $\Omega$ for EV.

After taking the contact resistance into account, we obtain for the maximum mobility of our samples: $\mu_{f,\text{XD}} \approx 5 \times 10^4$ cm$^2$/Vs, $\mu_{f,\text{EV*}} \approx 1.4 \times 10^5$ cm$^2$/Vs, $\mu_{f,\text{EV}} \approx 2.0 \times 10^5$ cm$^2$/Vs, and $\mu_{f,\text{EV2}} \approx 2.5 \times 10^5$ cm$^2$/Vs. These values are listed in Table I in the main paper. According to mobility, the quality of our samples follows the order EV2 > EV > EV* > XD. Fig. S2d illustrates the dependence of $\mu$ on the change density. The mobility on the hole side appears to be by factor of $3 - 5$ worse than on the electron side. We assign this difference to the presence of *pn* interfaces near the chromium-gold contacts which are known to have appreciable negative doping (4). The *pn* interfaces reduce the conductance and result in a lowered apparent mobility for holes.

### S3.1   Contact Resistance

The full calculation of conductance through graphene with contact transparency is quite a tedious task (4). Here we will describe in simple terms, how the contact

resistance can be viewed as an additive quantity on top of the graphene conductance which arises for graphene Corbino disk from the Dirac equation (*3*).

### S3.1.1 Conductance through a double barrier graphene system

To derive the total transmission through the sample with contact resistance, we follow the procedure of Ref. (*4*). First we note that usually phase coherence is more sensitive to disorder than reflection and transmission, and we expect that the former is destroyed but the latter is not affected by disorder (ballistic regime). This assumption is supported by the experimental results on the Fabry-Pérot resonances which are found to be weak. Furthermore, when destroying the Fabry-Pérot resonances fully by an applied bias, the overall conductance does not change appreciably. Hence, we conclude that incoherent treatment of transmission probabilities is well justified in our sample.

In general, for the case of incoherent tunneling through a symmetric graphene barrier with equal transmission $T$ for the left and right slopes, the total probability of transmission through the barrier is given by (*4*)

$$T_{\text{tot}} = \frac{1}{2T^{-1} - 1},$$ (S3)

If a finite contact resistance exists, then a finite transmission probability $0 \leq T_c \leq 1$ should be included in to the value of $T$ in Eq. (S3). By applying the incoherent addition model, we may write

$$T = \frac{T_c |t|^2}{T_c + |t|^2 - T_c |t|^2},$$ (S4)

where $|t|^2$ denotes the transmission probability through the graphene sample and $T_c$ gives the transmission at the contacts. This equation can be written in a more transparent form using resistances scaled by the quantum resistance $R_Q = \frac{h}{2e^2}$: $|t|^2 = R_{\text{ideal}}/R_Q$, $T_c = R_c/R_Q$, and $T = R/R_Q$. Then we obtain

$$\frac{R}{R_Q} = \frac{R_{\text{ideal}}}{R_Q} + \frac{R_c}{R_Q} - 1.$$ (S5)

By identifying $\frac{R_{\text{ideal}}}{R_Q} - 1$ as the scaled sample resistance (zero in the ballistic limit), we find direct additivity of the contact resistance with the ideal, calculable resistance of the sample.

### S3.1.2 Extracting Contact Resistance

Transport properties of a graphene Corbino disk were calculated in Ref. (*3*). The conductance is determined by

$$G = \frac{se^2}{h} \sum_{j=\frac{1}{2},\frac{3}{2},\frac{5}{2}\dots} T_j,$$ (S6)

where $s = 8$ and $T_j$ is the transmission coefficient of the $j$th channel. Transmission coefficient is given by

$$T_j = \frac{16}{\pi^2 k^2 R_1 R_2} \frac{1}{(\mathfrak{D}_j^+)^2 + (\mathfrak{D}_j^-)^2},$$ (S7)

with

$$\begin{aligned}
\mathfrak{D}_j^\pm = \Im \{ &H_{j-1/2}^{(1)}(kr_i) H_{j\mp1/2}^{(2)}(kr_o) \\
&\pm H_{j+1/2}^{(1)}(kr_i) H_{j\pm1/2}^{(2)}(kr_o) \},
\end{aligned}$$ (S8)

where $H_\nu^{(K)}(x)$ is the $K$th Hankel function and $k = \sqrt{\pi|n|}$ denotes the wave vector.

By using Eq. S6 one can generate theoretical curves corresponding to our devices. In Fig. S2e, the measured conductance is compared with three theoretical curves and the data are found to agree well when a contact resistance of 420 Ω is taken into account. Note that if we would linearize the dependence of $G$ vs. $n$ using $R_c$, a procedure which is expected be valid for long-range Coulomb scattering, then the contact resistance would grow from the above value. In our understanding, the estimated $R_C$ is mostly coming from the central leads where the multiple e-beam steps may create the lattice defects. Additionally, in our fabrication process, the development process for central leads has limitation to wash out the residual resists. These resists may forms the dominant charge density under the central metallic leads (*5*). Either or both phenomena could explain the origin of this contact resistance of 1 kΩ per micrometer, which is quite close to the experimental findings of Russo *et al.* (*6*).

### S3.2 Shubnikov de Haas Oscillations and Quantum Mobility

Samples were characterized at low magnetic fields around a few hundred mT using Shubnikov-de Haas oscillations (the well-defined quantum Hall states in our samples first develop around 0.5 T). These oscillations yield information about the

characteristic properties of charge carriers like mobility, quantum life time of carriers, and their effective mass. The diagonal resistance $R_{xx}$ oscillates periodically with inverse magnetic field, while temperature suppresses these oscillation due to enhanced scattering of carriers. The variation of resistance is given by equation (7):

$$\frac{R_{xx}}{R_0} = 1 + 4\frac{X}{\sinh(X)}\exp\left(-\frac{\pi}{\mu_q B}\right)\cos\left(\pi\frac{E_F}{\hbar\omega_c} - \theta\right), \qquad (S9)$$

where $R_0$ is the resistance at zero field, $X(T) = 2\pi^2 k_B T/\hbar\omega_c$ with cyclotron frequency $\omega_c = eB/m^*$, $E_F$ specifies the Fermi level, $\mu_q$ denotes the quantum mobility of charge carriers, and $\theta$ is a phase factor. Looking at resistance oscillations alone, Eq. S9 can be written as

$$\Delta R_{xx} = A\cdot\exp\left(-\frac{\pi}{\mu_q B}\right)\cdot\cos\left(\pi\cdot\frac{E_F}{\hbar\omega_c} - \theta\right), \qquad (S10)$$

with $A$ as a constant defined by $X(T)$.

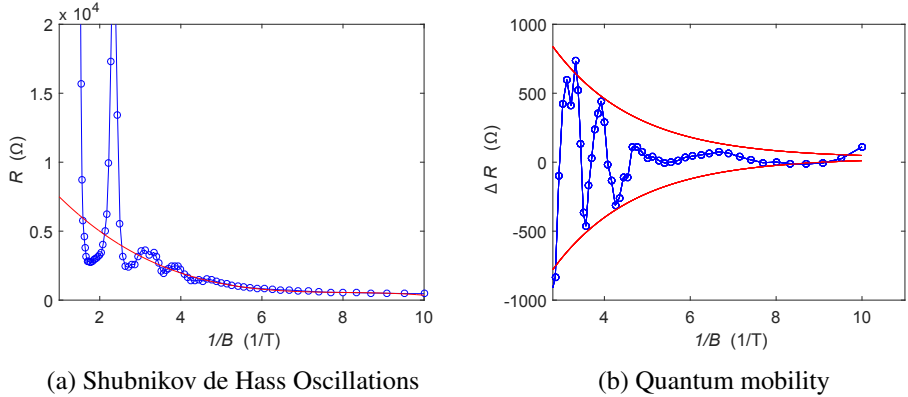

(a) Shubnikov de Hass Oscillations  (b) Quantum mobility

Figure S3: Quantum mobility calculation: a) Resistance of sample EV as a function of the inverse of the perpendicular magnetic field $1/B$ at $V_g = 10$ V. Shubnikov de Haas oscillations are visible around the base line indicated by the red curve. b) Shubnikov de Haas oscillations around the subtracted base line (red curve in Fig. S3a). The solid red curves represent envelope of the oscillations and they correspond to quantum mobility $\mu_q = 6 \times 10^4$ cm$^2$/Vs.

Fig. S3a displays the SdH oscillations seen in sample EV. The quantum mobility obtained from the fit amounts to $\mu_q \simeq 6 \times 10^4$ cm$^2$/Vs, which is by a factor of two smaller than the value of the field effect mobility in Table I of the main paper. This value for $\mu_q$ corresponds to the quantum scattering time $\tau_d \simeq 0.3$ ps

which yields for the Landau level width $\delta_{LL} = \hbar/2\tau_q \simeq 10$ K. In order to achieve a better analysis of the oscillations themselves, a base line marked by the red line in Fig. S3a was subtracted from $R$. These pure oscillations are displayed in Fig. S3b . Quantum mobility was extracted from the envelope curve of these oscillations, as indicated in Fig. S3b. In the case of sample EV, the quantum mobility is $\mu_q = 6 \times 10^4$ cm$^2$/Vs (red lines in Fig. S3b ).

## S4   Analysis of Transconductance Lines

The transconductance $g_m$ measurements were performed to identify fragile fractional fractional quantum Hall states in a well-controlled, indisputable manner. For $g_m(n, B)$ measurement, an ac excitation of the 2.7 mV$_{rms}$ is applied to the back gate on top of the dc voltage, and the sample is voltage biased with $V \sim 0.1 - 3$ mV. Both amplitude and phase of the current modulation were recorded using a lock-in amplifier. The gate modulation frequency was selected to correspond to the roll-off frequency of $RC$ time constant, governed by the sample resistance of a few M$\Omega$. Consequently, the change in $g_m$ was observed both in the phase angle and amplitude of the recorded signal. It turned out that, when using long integration times in our lock-in detection, the phase signal was easier to analyze for identification of FQH states than the amplitude signal. Hence, we found our analysis mostly on the phase of the transconductance signal.

Transconductance-based Landau fan type of diagram for sample XD is displayed in Fig. S4b. Compared with Fig. 2B in the main paper, this scan has more overlapping fringes, which tends to broaden the observed features and to smear the results of the correlation analysis. The number of visible features indicates a stronger disorder potential present in this device compared with sample EV, which is in agreement with the worse field-effect mobility listed in Table I for sample XD.

The FQH state signatures in $g_m$ are identified using correlation analysis, which is based on the work of Lee *et al.* (8). A line with a slope corresponding to a filling factor $\nu$, is drawn on the pixel map of $g_m(n, B)$ and the nearest pixels $(n_i, B_j)$ are selected. The values at selected points $(n_i, B_j)$ are transferred to vectors $A_k^w$ where the index $k$ covers the full line and the index $w$ keeps track of the vertical ($B$) coordinate of the drawn lines (see Fig. S4a). The cross-correlation $h(\nu)^w$ of pixels within a single vector $A_k^w$ is calculated using

$$h(\nu)^w = \frac{1}{S}\sum_{kl} A_k^w \cdot A_l^w (1 - \delta_{kl}),$$ (S11)

where the Kronecker delta $\delta_{kl}$ removes the auto-correlation part from the function and $S$ is the total number of pixel products in the summation over $k$ and $l$.

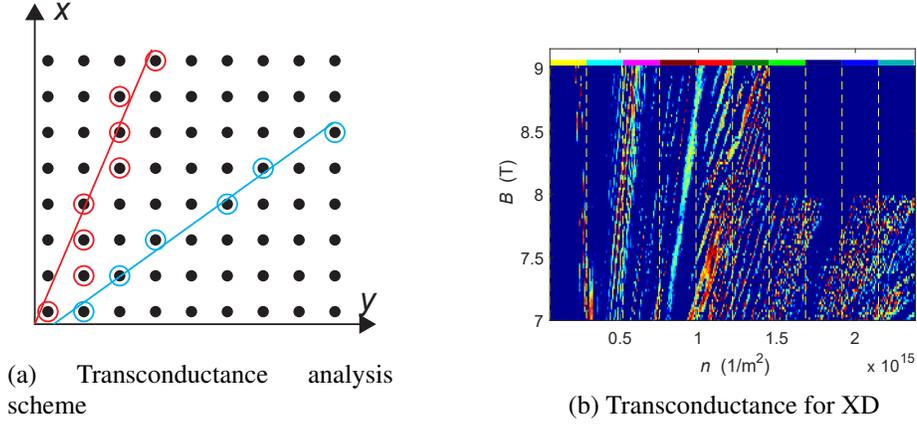

(a) Transconductance analysis scheme

(b) Transconductance for XD

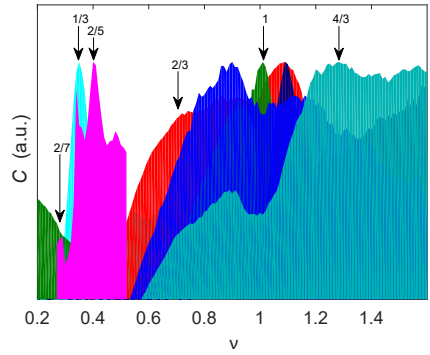

(c) Fractional states identification

Figure S4: Transconductance analysis: a) Vector $A_k^w$ with slope $\nu$ is formed by selecting a line $y = \nu x + w$ on the measured $n - B$ plane and taking the pixels closest to the line on each row (*i.e.* a particular field $B$) as illustrated in the figure. Blue and red lines correspond to filling factor slopes $\nu_1$ and $\nu_2$, respectively, and the black dots denote data pixels: the selected data points for correlation analysis are indicated by blue and red circles for these two filling factors. b) Measured transconductance phase data at high field $7 - 9$ T for sample XD (the blank upper right corner was not scanned in the experiment). The dashed yellow lines mark the boundaries of the sections in which the correlation function was evaluated. The color on top of each stripe corresponds to the analyzed histogram in Fig. S4c. c) Scaled correlation function $c(\nu)$ obtained from the data in Fig. S4b. The identified fractional states are marked by arrows; states marked in red could not be resolved in conductance measurements directly

10

The total unscaled correlation function is obtained by summing over all values of $w$: $h(v) = \sum_w h(v)^w$. In order to find dominant fractional slopes in the data, the cross-correlation was calculated for all $v \in \{0...1.5\}$ and histograms of $h(v)$ were plotted. A peak in the correlation histogram indicates presence of a dominating feature corresponding to the filling factor $v$. For a better inter-comparison of different analyzed regions, we scale our correlation histogram by the maximum correlation: $c(v) = h(v)/h(v)_{max}$.

The correlation map for sample XD is shown in Fig. S4c. As already mentioned above, the correlation peaks in this picture are broadened in comparison with Fig. 2C of the main paper. This is because the disorder potential in sample XD is larger than in EV, and disorder will in general suppress the observation of FQH states via life time broadening. Nevertheless, the transconductance measurement is able to bring about many of the fractional states in spite of the disorder.

The data in Fig. 2B in the main text display also clear fringes with slopes $v = 1$ and 2 overlaid in the middle of the picture on the bulk fringes. This co-existence of fringes representing different filling factors $v$ requires the local charge density $n_{loc} > eB/h$. Furthermore, these slopes are by far off from the possible slopes facilitated by the width of the residual charge distribution based on $\log G(V_g)$ analysis. Hence, we conclude that $v = 1$ and 2 fringes are located at the boundary regions near the metallic contacts. Due to absence of any polarizing medium close to suspended graphene, the screening of charge near metallic contacts is rather weak. Consequently, strong doping from a metallic contact can extend appreciably in to the suspended graphene near the contact (*9*). The amount of edge doping depends on the details of the metallic contacts, which may be in the range of $E_F \sim -100$ meV (*4*). Such a doping corresponds to a charge density of $n_c = 7.2 \times 10^{11}$ cm$^{-2}$ > $eB/h$ under the contact metal, which is sufficient for accommodating QH states with $v = 2$.

# S5 Magnetoconductance

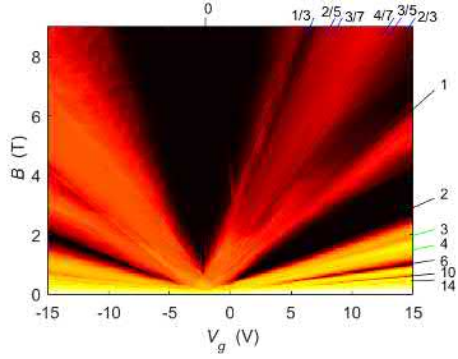

(a) Magnetoconductance $\log(G_d)$ Landau fan diagram

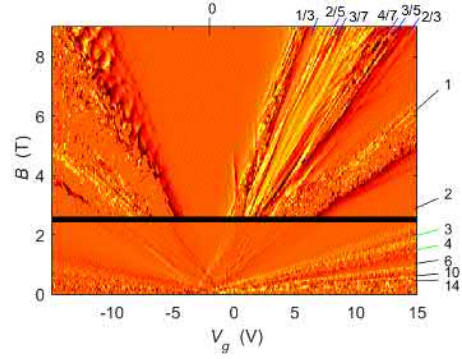

(b) Differentiated magnetocoductance $\delta \log(G_d)/\delta n$ Landau fan diagram

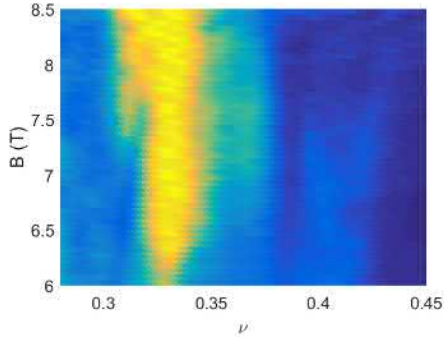

(c) Composite Fractional states identification

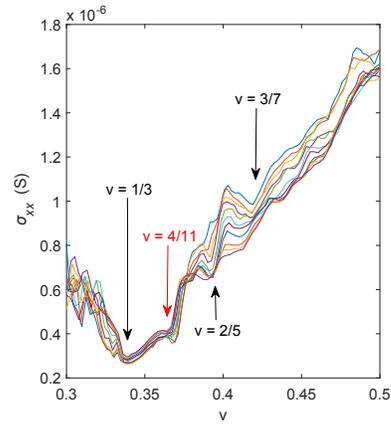

(d) Differential conductivity $\sigma_{xx}$

Figure S5: Magnetoconductance analysis: a) The magnetoconductance ($\log(G_d)$) measurement of sample EV showing broken symmetry states $\nu = \{0, 1, 3, 4\}$. Here charge density ranges over $n = -1.1 \times 10^{11}... + 1.5 \times 10^{11}$ cm$^{-2}$ and $G_d = dI/dV|_{V=0}$. b) The magnetoconductance ($\log(G_d)$) measurement of Fig. S5a differentiated with respect to charge density. The high field region, where the scale is by a factor of 100 smaller than the low-$B$ region, is separated using a horizontal black stripe. c) Inverse magnetoconductance $G_d^{-1}$ on the plane spanned by the filling factor $\nu$ and magnetic field $B$ measured for sample EV2. Interacting CF states $4/13 = 0.308$ and $4/11 = 0.363$ are observed next to $\nu = 1/3$ state above critical field values of 6.8 and 6.3 T, respectively. d) The differential conductivity $\sigma_{xx} = G_d \frac{\ln(r_o/r_i)}{2\pi}$ vs. the filling factor $\nu$ for sample EV, measured at magnetic fields $B = 6 - 7.5$ T.

The Landau fan diagram measured on sample EV (see Table I in the main paper) is illustrated in Fig. S5a. In addition to several basic integer Landau levels (2, 6, 10, 14) and broken symmetry states (1, 3, 4), several fractional states such as 1/3, 2/5, 3/7, 4/7, 3/5, and 2/3 are visible in the conductance. However, the levels are visible on the electron side but not clearly on the hole side [1]. Consequently, our magnetoconductance work concentrates on the FQH states on the electron side.

Fig. S5b illustrates the data of Fig. S5a after differentiation with respect to charge density. By comparing these two figures for $G_d$ and $d\log(G_d)/dn$, it becomes immediately clear that the differentiated data provides a much better resolution for identifying weak fractional states, including those of interacting composite fermions (CF). This is well evidenced by the data displayed in Fig. 2A in the main paper.

The interacting composite fermion states are visible also in our data on the differential inverse magnetoconductance $G_d^{-1} = dV/dI|_{V=0}$, which is depicted in Fig. 3A in the main paper for sample EV2. The data display two interacting CF states at $v = 4/11$ and $v = 4/13$, located on opposite sides of 1/3 state. The $G_d^{-1}$ data on sample EV2 is replotted in Fig. S5c as a surface plot on $v$ vs. B plane. We note that, in this plot, all peaks assigned to FQH states are well vertical and they do not shift with field over the range $B = 6 - 8.5$ T.

The data in Fig. S5c clearly indicates the independence of the peaks at $v = 1/3$ and at $v = 4/11$: the peak at $v = 4/11$ appears narrower and it emerges at clearly different field value as the peak at $v = 1/3$. The same conclusions apply for the comparison of $v = 1/3$ with the peak at $v = 4/13$. The signature at $v = 4/11$ was also confirmed in mixing current experiments probing non-linear conductance at higher frequencies around 20 MHz where the hopping conductance becomes stronger.

In addition to sample EV2, the interacting CF states 4/11 and 4/13 are well visible in the $G_d = dI/dV$ vs. $v$ traces of sample EV measured at magnetic fields $6 - 7.5$ T. These data are displayed in Fig. S5d. In these data, the state 4/11 is better visible than 4/13, which is to be expected since the former fraction is related to $2\Phi_0$ composite fermion series while the latter belongs to interacting $4\Phi_0$ CF particles.

---

[1] Asymmetrical magnetoconductance *i.e.* absence of fractional quantum Hall states on *p* side may be caused by the presence of snake states at the *pn* interface, which makes the nature of the particles at this interface to oscillate between an electron and a hole. Such conversion oscillation, when coupled to localized states of hole carriers, is quite likely to lead to broadening of the localized Landau levels.

## S6   Basic transport processes

The IV characteristics of our Corbino samples contain contributions from many different transport processes: 1) quantum tunneling, 2) thermally assisted quantum tunneling, 3) thermal activation, 4) phonon assisted hopping, 5) electron crystal sliding *a la* charge density wave. All these processes leave their characteristics in to the IV curves. Furthermore, we have measured IV characteristics under rf-irradiation (see Sect. S9) at which the transport is influenced by photon assisted hopping, 7). Here we will discuss only those topics which are most relevant for the explanation of our data, namely items 3 and 7 above. However, one should keep in mind that processes in items 2 and 3 are intimately connected, and parameters for thermal activation may become adjusted by quantum tunneling near the top of the energy barrier.

### S6.1   Thermal activation

The conductance in the bulk of a quantum Hall conductor depends intimately on the potential landscape of the state of the system. The potential landscape leads to localized states as illustrated in Fig. S6a. Theoretical models in the case of a long-range random disorder potential predict thermally activated conductivity,

$$\sigma = A\exp(-U_s/2k_BT), \tag{S12}$$

where the activation energy $U_s \simeq C\Delta_\nu$ is proportional to the gap $\Delta_\nu$ in FQH states with $C \lesssim 1$, and the prefactor $A \simeq (e^*)^2/h$ is governed by effective charge $e^*$ of charge carriers (*10, 11*); in thermal activation, the effective charge is related to the filling factor $e^* = \nu e$ (*10, 11*), whereas in quantum tunneling regime it may be either $\nu e$ or $e$ (*12*). The suppression factor $C$ of $\Delta_\nu$ in the activation formula depends on the significance of quantum tunneling near the top of the barrier. In a similar approximative approach, the conductivity in the quantum tunneling regime can be expressed as $\sigma = A\exp(-U_s/2k_BT_{cr})$, where $T_{cr}$ denotes the cross-over temperature between thermal activation and quantum tunneling. The cross-over temperature is difficult to determine theoretically but it can be estimated from the saturation temperature of $\sigma(T)$ scans, which in our experiments varies in the range $T = 1 - 0.5$ K.

Our conductance at high $T$ agrees with the thermal activation process (*10, 11*) and, in this regime, the data can be employed to obtain estimates of the excitation energy gaps of the FQH states by assuming $\Delta_\nu = U_s$. By determining the activation energy across $\nu = 1/2$ and assuming linear dependence for $\Delta_\nu$ vs $\nu$, we obtain approximatively 0.2 meV for the composite fermion LL width. This can be regarded as a systematic uncertainty (offset) for the energy gaps quoted in our

paper for FQH states. Fig. S6b displays conductance data for the state $\nu = 4/3$ for the sample EV2. The data indicate thermally activated conduction at temperatures above 1 K. By using Eq. S12, we obtain for the energy gap $\Delta_{4/3} = 0.7$ K. At low temperatures, especially at low filling factors, the role of thermal activation becomes weaker, and the conductance and IV characteristics display a more complex dependence on charge density as will be discussed in Sect S7.

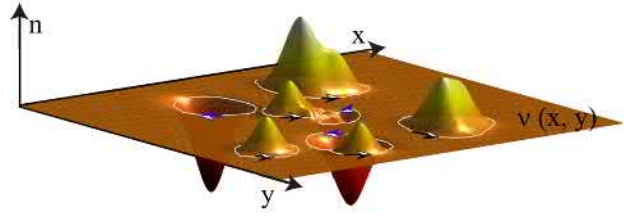

(a) Localized states

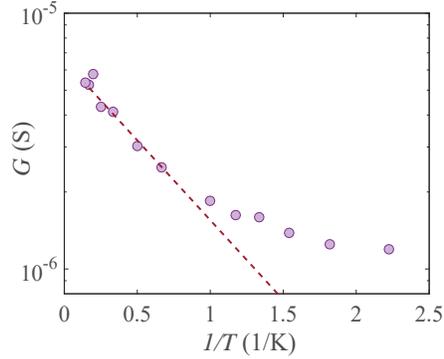

(b) Activation gap energy $\nu = 4/3$

Figure S6: Transport process: a) Schematic picture of localized states with variation in charge density marked in green (enhanced density) and red (reduced density). The localized states are embedded within the incompressible FQH liquid with constant electron density (brown plane). Landau levels (LLs) are bending upwards when approaching the red regions while they bend down around the green bumps. Consequently, skipping orbits of edge state electrons reside outside and inside the borders of the low and high density regions, respectively. White-blue arrows mark the directions of the currents of these localized edge states. Electric transport at low temperatures takes place via quantum tunneling between the localized states through a barrier due to the energy gap $\Delta_\nu$ of the incompressible FQH region. In case of considerable variation of electron density (*i.e.* the charge dips/hillocks reaching the next FQH charge density plane), there will be additional localized fractional states within such regions. Since these states are localized within a bounded region with an edge state, their effect on charge transport can be neglected. b) Logarithm of the zero-bias conductance $G$ vs. inverse temperature $1/T$ for sample EV2 at magnetic field $B = 4$ T. The dashed line corresponds to an energy gap of $\Delta_{4/3} = 0.7$ K

## S6.2   Photon assisted hopping

Hopping conduction in a disordered conductor having localized states is enhanced by microwave irradiation. This problem has been reviewed e.g. by Efros and Shklovskii (*13*). We will deal with this issue only qualitatively, and our discussion is inspired by the review of Galperin and Gurevich (*14*).

It was first shown in Ref. (*15*) that, under microwave irradiation, the hopping conduction across localized states is determined by so called close pairs. Each such pair consists of one localized state which is occupied by an electron and one localized state which is empty, the distance between them being much smaller than the average distance between localized states given by $\langle r \rangle = (\frac{4\pi^2}{3} N_A)^{-1/2}$ where $N_A$ is the density of localized states. Due to absorption, electron occupancies in the pairs are changed, which is equivalent to an enhancement of the average energy of bound electrons. Thus, microwave absorption leads effectively to heating of the system of localized states. Below we determine the conductivity $\sigma(\omega)$ that is relevant for the absorption of irradiation in the presence of localized states in a three dimensional system. As usual in hopping conductance, we don't expect any qualitative change between different dimensionalities, only a change in the exponent $\alpha$ of the conductivity $\sigma \propto \exp(\frac{E_0}{E}^\alpha)$ in terms of characteristic energy. This difference in the exponent $\alpha$ is irrelevant for the following discussion.

We are interested in the regime in which $e^2/(4\pi\varepsilon\langle r \rangle) >> \hbar\omega, k_B T$, i.e. the charging energy of the localized state is the largest energy scale. In the resonant case ($\hbar\omega$ matches the energy difference between the levels in question), the Ohmic microwave conductivity is given by

$$\sigma_0^{res}(\omega) = \frac{\pi^2}{3} ag^2 \frac{e^4}{\varepsilon} \omega r_\omega^3 \tanh\left(\frac{\hbar\omega}{k_B T}\right).$$ (S13)

Here $g$ denotes the single particle density of states, $a$ corresponds to the radius of the localized state, and

$$r_\omega = a \log\left(\frac{\Lambda_0}{\hbar\omega}\right)$$ (S14)

specifies the minimum distance between pairs with energy separation $\hbar\omega$; $\Lambda_0$ denotes the off-diagonal elements in the tunnel coupling matrix. As $r_\omega$ does not depend on temperature, the resonant conductance $\sigma_0^{res}(\omega)$ decreases with increasing $T$. Similarly, a strong drive will result in more equalized populations, which leads to decrease of absorption. In both cases of reduces absorption, relaxation of the close pairs becomes of major importance.

In the non-resonant (relaxational) case with fast relaxation $\omega\tau_1 << 1$, where

$1/\tau_1$ describes the relaxation rate due to the environment, the conductance becomes

$$\sigma_0^{rel}(\omega) \approx ag^2 \frac{e^4}{\varepsilon} \omega r_e^3.$$ (S15)

In our experiments based on mixing current detection of the FQH states, the relevant frequency is around 20 MHz (see Sect S10). We expect that the localized states will remain in the incoherent in this regime (i.e. relaxation time $\tau_1 < 50$ ns). In this case, the relevant pairs have an energy separation $E \sim k_B T$ and a life time $\tau_{min} \sim \omega$, and the characteristic minimum distance between pairs becomes

$$r_e = a \log\left(\frac{\Lambda_0}{k_B T [\omega \tau_{\min}(T)]^{1/2}}\right).$$ (S16)

By using electron-phonon supercollision scattering rate near the Dirac point (*16*), we observe that $r_e$ increases with $T$, which leads to growth of relaxational conductance $\sigma_0^{rel}(\omega)$ with temperature.

To sum up this section, hopping conductance grows with $\omega$ and, in magnetic fields where $\ell_B \sim \langle r \rangle$, we may have $\sigma(\omega) >> \sigma(0)$. In addition, $\sigma_0^{res}(\omega)$ due to resonant conductance is expected to dominated by non-resonant relaxation $\sigma_0^{rel}(\omega)$, which will enhance both as a function of $T$ and microwave power in the limit of $\omega \tau_1 << 1$. This microwave absorption will result in heating of graphene that, in turn, will govern the change in low frequency (or DC) conductance in the measurements on our sample.

## S7  *IV*-characteristics

Our *IV* characteristics at low filling factor ($\nu = 0.16...0.33$) are illustrated in Fig. S7a for sample EV. At $\nu = 0.33$, the *IV* curve displays first a linear part that extends nearly up to $0.5 - 0.7$ mV, above which there is a strong increase in current in a power-law-like fashion as $V^\alpha$. Such an *IV* curve can be understood by transport via quantum tunneling along a percolating path of several localized states where the non-linear regime would be a sign of competing order of correlated tunneling of chiral edge states (*17, 18*).

At higher temperatures, this conductance at $\nu = 0.33$ becomes weakly temperature dependent, which is an indication of thermally assisted quantum tunneling. The combination of thermal activation and quantum tunneling leads to a lowered effective gap as determined from the temperature dependence of conductance (*11*). At $V > V_T \simeq 1$ meV, the *IV* characteristics becomes a power law $\nu^\alpha$ with an exponent of $\alpha = 3 - 4$. We assign this to edge-to-edge tunneling of fractional charge

between localized states along the percolating path[2].

At filling factor $\nu = 0.20$, even the low voltage part appears as non-linear and it is difficult to resolve any Ohmic part in the $IV$ curve, This behavior is identified as a cross-over regime between FQH and Wigner crystal order: Here pinning of the crystallites in the Wigner phase is not fully developed and linear quantum tunneling and non-linear sliding processes of charge are competing, resulting in an $IV$ curve that is difficult to classify.

At $\nu = 0.16$ there is a linear regime that is smaller than at $\nu = 0.33$. The behavior of our non-linear $IV$ characteristics for sample XD is illustrated in Fig. 4B in the main paper: the data at $\nu = 0.16$ are seen to display exponential increase in current between $V \sim 0.1 - 2.0$ mV. Here we display the $IV$ characteristics of sample EV2 in Fig. S7b at filling factor $\nu = 0.15$. The data indicate similar thermal depinning of Wigner crystal as is seen in the XD sample. Eq. 1 of the main paper has been used to fit the data using parameters $e^* = e$, $T = 0.2$ K, $\Delta = 170$ $\mu$V, and $N = 5$. These values are consistent with results on the XD sample as well as with the results of the same sample under rf irradiation (see below).

Fig. S7c displays the logarithm of total conductance $G_{tot} = I/V$ on the $\nu - V$ plane for sample XD. The linear low-bias structure of the $IV$ curves at $\nu > 0.2$ is seen as constant conductance regions below the power-law regime. Above the power-law regime, $G_{tot}$ gradually approaches the high-bias slope $dI/dV \sim 100$ k$\Omega^{-1}$. Fig. S7c indicates that conductance is strongly suppressed between $\nu = 0.12 - 0.20$, but good conductance re-emerges again at $\nu = 0.10 - 0.12$. For sample EV2, the Wigner crystal state found over a slightly smaller regime covering $\nu = 0.15 - 0.18$.

The $IV$ characteristics of EV2 in the re-entrant good conductance regime at filling factor $\nu = 0.14$ is illustrated in Fig. S7d. The behavior at $\nu = 0.14$ resembles that of $IV$ curves at $\nu = 1/3$, where the initial linear behavior changes to a power law with an exponent $\alpha = 3 - 4$, but now with an exponent $\alpha \simeq 7$. The shape of the $IV$ curve hints towards re-entrance of FQH states, but we are not able to identify any peak in the gate sweeps, which would make this conjecture more plausible. Nevertheless, as displayed in Fig. 3B in the main paper, we have determined the energy gap using $G(T)$ measurements and we can state that this intermediate conductance state is gapped. At smaller charge density, for example at $\nu = 0.12$, the $IV$ curve displays a larger gap than in the Wigner crystal regime, and the behavior gradually develops towards Zener tunneling $IV$ curves.

---

[2]Our exponent for tunneling between localized states above thermal energy is approximately twice that observed for metal-to-edge tunneling, see Jolad (*19*)

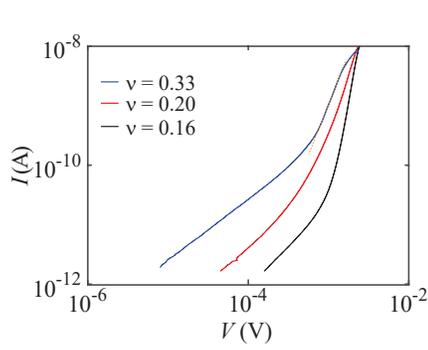

(a) IV curves at low charge densities

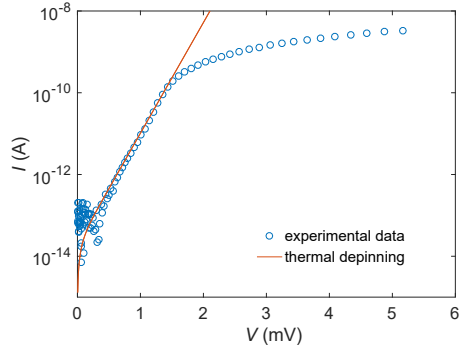

(b) IV characteristics of Wigner crystal

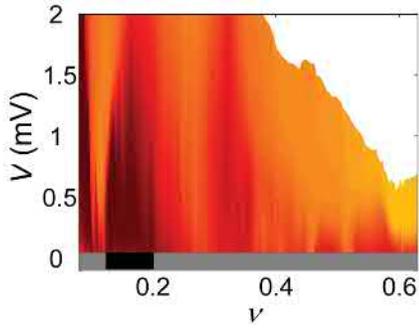

(c) Density plot of IV's

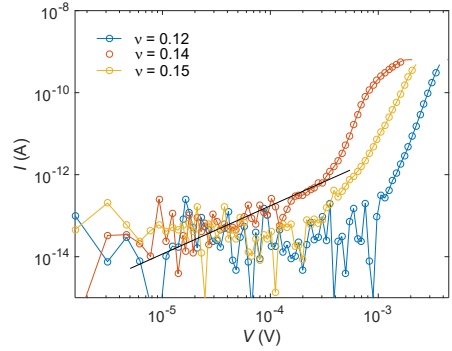

(d) IV analysis of re-entrant Fermi liquid

Figure S7: *IV* characterization : a) *IV* curves on log-log scale measured for sample EV at $B = 9$ T for fillings factors $\nu = 0.33$ (blue), 0.20 (red), and 0.16 (black). The fitted dashed line to $\nu = 0.33$ data in the non-linear regime indicates a power law with an exponent of $\alpha \sim 3.4$. b) *IV* curve of sample EV2 on semi-log scale measured at $\nu = 0.15$. The solid line indicates the exponential growth obtained from the thermal depinning model (Eq. 1 in the main paper) using parameters $e^* = e$, $T = 0.2$ K, $\Delta = 170$ $\mu$V, and $N = 5$. c) Logarithm of conductance $G_{tot} = I/V$ measured for sample XD as a function of bias voltage $V = 0 \ldots 2$ mV over filling factors $\nu = 0.05 - 0.65$ at $B = 9$ T; orange color indicates high conductance and dark red shows reduced conductance (range spans $2 \cdot 10^{-5} - 1 \cdot 10^{-8}$ S). The Wigner crystal regime $\nu = 0.12 - 0.20$ is indicated by the black bar on the grey bottom stripe. d) *IV* curves of sample EV2 measured at slightly smaller electron density of the Wigner crystal behavior at $\nu = 0.15$. The solid black line denotes linear *IV* characteristics.

## S8   Resonance modes in pinned Wigner crystal

The nature of pinning in disorder Wigner crystal depends strongly on the characteristics of the disorder. In suspended graphene membrane there are two quite different disorder elements present. First, there are membrane ripples (dimples), which can be either static or alternatively dynamic due to thermal fluctuations. Second, residual charge density is related to charged impurity disorder, most likely caused by residues of the lithography process, which are present in spite of the current cleaning. Static ripples can be modeled by the interfacial disorder model of Ref. (*20*), which relates the pinning potential of the Wigner crystal to the depth and width of the interfacial dimples. The pinning potential caused by charged impurities in clean graphene has a rather long-scale variation so that the particle size (magnetic length $l_B$) is smaller than the correlation length of the disorder $\xi$. Such a pinning case has been treated for example in Ref. (*21*). We expect that both pinning mechanisms will contribute to the pinning potential and contribute to the pinning frequency. By assuming that the disorder contributions are independent, we obtain the pinning frequency $\omega_p$ as a combination of these two processes: $\omega_p^2 = \omega_q^2 + \omega_r^2$ where $\omega_q$ and $\omega_r$ refer to charged impurities and interfacial disorder, respectively. Below we will outline the basic pinning energy considerations in both disorder cases.

### S8.1   Pinning by charged impurities

A Wigner solid is pinned in the presence of disorders, which results in breaking of the long-range order of the crystal. Consequently, the crystal splits into coherent, independent crystallites, and a gap opens up in the transverse phonon modes of the Wigner lattice. In our graphene case, the correlation length $\xi > l_B$, and the particle locally sees a smooth potential, for which the particle size $l_B$ is not directly relevant. Under these conditions, the crystallite size $L_c$ over which collective pinning takes place is independent of the magnetic field. The crystallite size specifies the collective-pinning-mode frequency of the Wigner solid. One obtains for the pinning frequency in the simplest approximation

$$\omega_p^0(B) = (2\pi^2)^{-\frac{1}{6}} \frac{cR_a^{-2}}{nB} (\frac{a}{\xi})^6, \tag{S17}$$

where $c$ is the shear modulus of the crystal, $a$ is the lattice constant of the Wigner crystal, and $R_a = ca^2/n_0\Delta$ is the length scale at which the fluctuation of the electrons in the Wigner crystal starts to be on the order of $a$; $n_0$ denotes the average charge density. Consequently, this kind of pinning sites yield a contribution to pinning frequency which decreases as $B^{-1}$. Using realistic values for $\Delta$ and $\xi$,

we obtain pinning frequencies in the GHz range. The relation of the pinning frequency $f_p$ to the crystallite size is given by $L_c = (2\pi c/n_i e B f_p)^{1/2}$, where $c$ denotes the shear modulus of the Wigner crystal. By assuming the classical form for the shear modulus in a 2D solid, $c = 0.245 e^2 n_i^{3/2}/4\pi\varepsilon_0\varepsilon_g$, and using $\varepsilon_g \sim 3$, we obtain for the domain size of the Wigner crystal $L_c = 0.70$ $\mu$m at $B = 9$ T. This is approximately half of our sample size $r_o - r_i = 1.15$ $\mu$m.

Commonly, charged impurities are viewed as a paradigm for a strongly pinned system (*22*), where certain locations of the lattice are basically fixed and, as such, eliminated as degrees of freedom. However, it is argued in Ref. (*20*) that one may consider a strongly pinned Wigner crystal as a weakly pinned crystal with point defects. If so, then the strongly pinned Wigner crystal for the purposes of AC conductance is equivalent to a weakly pinned, defective WC. Such arguments provide further support for the general applicability of the weak pinning formulas used in the main paper.

## S8.2 Pinning by ripples of membrane

Here we assume that the dimple lattice of suspended graphene can be approximated by the model of interfacial pits of depth $\Delta z$ and width $w$ developed in Ref. (*20*). The scanning tunneling microscope results of Ref. (*23*) indicate, for a typical current-cleaned graphene sample such as ours, an amplitude of corrugation of $Z = 0.4$ nm with a radius of curvature $R = 1.5$ nm (i.e. wave length $\sim 5$ nm). Hence, we may assume $\Delta z = 0.4$ nm and $w = 3$ nm. In our sample, the ratio of $\Delta z/\varepsilon$ is the nearly the same as that used in the evaluations in Ref. (*20*) (both $\varepsilon_g = 3$ and $\Delta z \sim 0.4$ nm are by a factor of $\sim 3$ smaller in graphene when compared with GaAs heterostructures). The potential energy gained by placing an electron in to dimple that is much larger than the magnetic length is given by $\Delta E = 2\pi n_0 e^2 \Delta z/(4\pi\varepsilon_0\varepsilon_g) \approx 7$ K in temperature units. Since a typical dimple is generally be much smaller than the magnetic length, the pinning energy of an electron trapped in a single dimple will be on order of $\Delta E w^2/l_B^2 \ll \Delta E$.

As in Ref. (*20*), we will assume the average distance between dimples $1/\sqrt{n_i}$ is smaller than the lattice spacing of electrons $a$ but larger than the magnetic length $l_B$. This means that each unit cell of the Wigner crystal contains several dimples on to which an electron could be trapped. We will ignore the spread in the size distribution of dimples and assume that their distribution can be approximated by single size and uniform density ($> 10^{11}$ cm$^{-2}$), the latter of which will form a fitting parameter, to be determined self-consistently in the model.

The average number of dimples occupied by an electron in a single domain of size $L_c$ if placed at a random location is given by $N_{pin} = \pi n_0 L_c^2 l_B^2 n_i$. Consequently,

the energy per electron gained from the disorder potential $u_p$ is

$$u_p \approx \Delta E \frac{w^2}{l_B L_c} \left[ \frac{\pi n_i}{n_0} \right]^{1/2} .$$ (S18)

By assuming purely transverse distortions in the lattice, we obtain an estimate for the distortion energy per electron

$$u_d \approx \frac{c}{n_i L_c^2 n_0}$$

where $c$ is the shear modulus of the lattice. Minimizing the total energy $u_d - u_p$ with respect to $L_c$, we obtain for the optimum domain size

$$L_c = \frac{2c l_B}{\Delta E w^2 (\pi n_i^3 n_0)^{1/2}} .$$ (S19)

The resulting pinning energy per particle is found by substituting this value of $L_c$ into our expression for $u_p$ in Eq. S18. Using standard lattice parameters for the shear modulus, the resulting pinning energy per particle can be written as

$$u_p \approx 5.24 \frac{4 \pi \varepsilon a^3 \Delta E^2 w^4 n_i^2}{e^2 l_B^2} .$$ (S20)

In the simplest approximation, the pinning frequency is given by the average binding energy per site, *i.e.*, $\hbar \omega_{pin} \approx u_p$. Note that within this approximation, $\omega_{pin} \propto l_B^{-2} \propto B$, which leads to a pinning frequency that *increases* with magnetic field. Using $n_i \approx 2.5 \times 10^{11}$ cm$^{-2}$, $l_B \approx 9$ nm, $w = 3$ nm, and $\Delta E / k_B = 7$ K, the pinning frequency becomes $\sim 1.5$ GHz. This value is strongly dependent on $\Delta E$ or the dimple size $w$, which could even be larger than 3 nm (*23*).

To sum up this section, we have two independent pinning potentials for the electrons in the Wigner phase. Charged impurities lead to pinning frequency $\omega_q \propto B^{-1}$ while dimples yield $\omega_r \propto B$. These pinning frequencies are added as independent contributions so that they yield the total pinning frequency $\omega_p^2 = \omega_q^2 + \omega_r^2$. Since the pinning frequency does not display much magnetic field dependence, $|df_p/dB| < 100$ MHz/T, we conclude that both contributions are approximately of equal magnitude. Consequently, the pinning frequency used in deducing the crystallite size, is only approximate and it yields an upper limit for the crystallite size.

## S9    Wigner crystal under rf irradiation

Commonly, microwave conductance measurements have been conducted on GaAs 2DEG by capacitively coupling RF power to the 2DEG and measuring its transmis-

sion through the sample (at high frequency). FQHSs, Wigner phases, and CDW-phases have been studied with this technique (*24–28*). Similar to earlier works, we have also probed the Wigner crystal by microwave spectroscopy. Our measurement method has two distinct features; first, the RF power is directly coupled to the Wigner crystal via the inner contact of the Corbino geometry and, second, the transport across the Wigner crystal is measured at DC. Irrespective of these differences compared with earlier studies in 2DEG, the essence of our incompressibility measurements of the Wigner crystal under RF radiation will be the same.

A schematic of the measurement circuit is depicted in Fig. S8. We employed bias-T circuits (Minicircuits) at the mixing chamber plate in order to apply simultaneously DC and AC bias across the sample. DC (or low-frequency) conductance was measured in voltage bias configuration through inductors while microwaves were coupled via capacitors to the sample. In these conductance measurements, a DC bias of $V = 50\ \mu V$ was applied to the inner contact while the resulting current at the outer contact was tracked using a Stanford SR570 current (transimpedance) preamplifier operating at a gain of $10^{10}$ V/A followed by a HP 34410A multimeter. The DC bias was provided using a HP 33120A generator connected through a 1:2000 voltage divider.

Anritsu MG3692B RF-generator was used to generate the applied RF irradiation at frequencies $f = 10$ MHz - 10 GHz. The RF-input line had an attenuation of 50 dB at $f = 1$ GHz, with additional $\sim 10$ dB per decade in frequency (i.e. attenuation was roughly 40 dB at 100 MHz, and 60 dB at 10 GHz). The RF-signal was fed to the inner contact through the bias-T, while the RF-line was terminated using a 50 $\Omega$ resistor at the outer contact.

There should always be hopping conductance of Efros-Shklovskii type, even in the case of a disordered Wigner crystal (*29*). As discussed in some detail in Sect. S6.1, this conduction is mediated by resonant pairs of localized states, the absorption of which grows with enhanced frequency. Since there is a large spread of energy differences among the resonant pairs in our sample, we expect to have an increase of absorption over a large band of irradiation frequencies. This is apparent in our experiments at large irradiation power as displayed in Fig. S9 for the strongly insulating regime at the filling factor $\nu = 0.12$. The RF absorption leads to increase of conductance via heating, and this enhancement in $G$ is seen to increase monotonically with frequency in Fig. S9.

At high irradiation power, Joule heating $P = \sigma(\omega)v_{ac}^2$ becomes relevant, and it leads to an increase of temperature and conductance in the sample. At low irradiation powers (around $-70\ldots-65$ dBm), however, only one strong peak in DC conductance around 3 GHz is observed (see Fig. 4A in the main paper). The resonant frequency coincides well with estimates for a resonance in a pinned, disordered Wigner crystal.

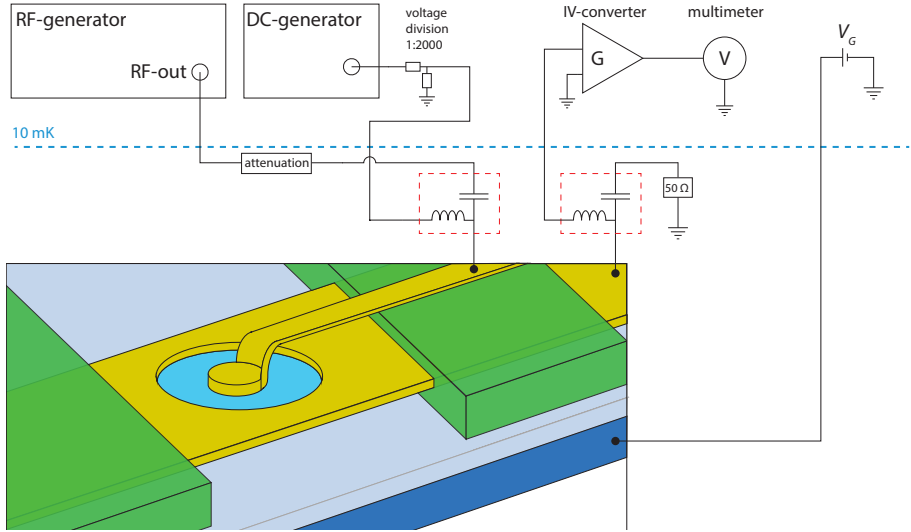

Figure S8: rf measurement setup: Schematic of the setup for conductance measurements under RF irradiation. Cryostat details and low pass filters in the DC-lines have been omitted from the picture for clarity.

The amplitude of this conductance resonance with respect to a broad background current enhancement grows with temperature. We assign this behaviour to a combined action of activated conductance and enhanced Joule heating by the microwave irradiation in the Wigner crystal. The Joule heating $P = \sigma(\omega)v_{ac}^2$ of the electron system is strongly frequency dependent due to the resonant behavior of $\sigma(\omega)$ at pinning frequency $\omega_p$. At the resonance, increase in temperature leads to an increase in observed conductance, and the change in conductance per unit Kelvin ($\Delta T$) will grow with increasing temperature due to increased microwave absorption. The conductance enhancement will continue till the Wigner crystal melts and the resonance absorption at $\sigma(\omega_p)$ is lost. This scenario supports the identification of the maximum in Fig. 4C of the main paper as the melting temperature of the Wigner crystal. Similar identification of the Wigner crystal melting has been employed in Ref. (*30*). For reference, we also show $G(T)$ without rf irradiation in Fig. S10. This plot on linear scale indicates beginning of melting at $T_{mb} = 1.4$ K, which is lower but consistent with the melting temperature $T_m = 1.7$ K obtained from rf resonance measurements.

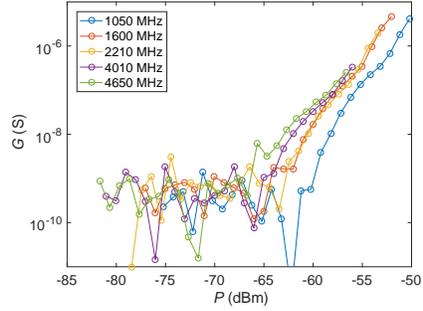

Figure S9: rf conductance analysis: Conductance as a function of input RF power at the Corbino disk measured in the frequency range $1 - 5$ GHz for EV2 at $\nu = 0.12$ (strongly gapped regime). The applied DC-bias was $V = 50\,\mu$V. A clear increase in the DC conductance $G = I/V$ is observed at power levels exceeding $\sim -65$ dBm. The power level used in the measurement of Fig. 4A in the main paper corresponds to $-68$ dBm in this graph.

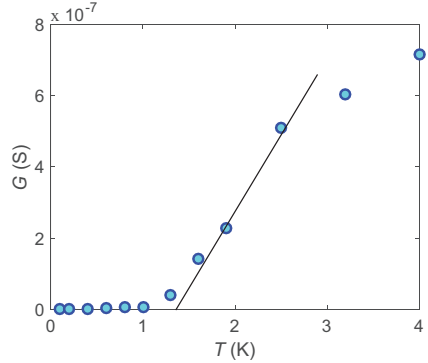

Figure S10: Melting of Wigner crystal: Temperature dependence of conductance $G(T)$ for sample EV2 on linear scale at measured at $\nu = 0.15$. The line marks linear growth, using which one obtains approximate melting temperature of $T_{mb} = 1.4$ K for the Wigner crystal phase.

## S10 FQH states under rf irradiation

In addition of studies of the FQH states under GHz irradiation, we have studied them at MHz frequencies using mixing currents facilitated by mechanical motion (*31*). The mechanical motion with amplitude $z$ yields a mixing current via differential conductance $G_d(\omega)$ as

$$I_{mix} \propto \frac{\partial G_d(\omega)}{\partial z} \left| \frac{\partial \mathrm{Re}(z)}{\partial f} \right|$$

when the sample is biased using a frequency modulated signal $V^{FM}(t) = V^{AC}(\cos(2\pi ft + (f_A/f_L)\sin(2\pi f_L t))$. Here $V^{AC}$ is the amplitude of the carrier frequency signal, $f_L$ is the frequency modulation, and $f_A$ defines the modulation depth. In the Wigner crystal and FQH transport regimes, the hopping conduction is enhanced by operation frequency so that the mixing current method may be more sensitive than the low frequency or DC conductance measurement. We have carefully measured the region around $v = 1/3$ and our experiments confirm the finding of the fraction state of $v = 4/11$ observed in the low frequency measurements.



\end{document}